\documentclass[prd,11pt,a4paper]{article}
\pdfoutput=1
\usepackage{jheppub}
\usepackage{rotating}
\usepackage{array}
\usepackage{amsmath,amssymb,amstext}
\usepackage[normalem]{ulem}
\usepackage{slashed}
\usepackage{booktabs}
\usepackage[pdftex,table,dvipsnames]{xcolor}
\usepackage{multirow}
\usepackage{url}
\usepackage{color}
\usepackage{xspace}
\usepackage{comment}
\usepackage{enumitem}
\usepackage{subfig}
\usepackage{siunitx}

\usepackage{multirow}
\newcommand{\Cathode}{CATHODE}

\preprint{P3H-24-077, TTK-24-45}

\title{Accurate and robust methods for direct background estimation in resonant anomaly detection}

\author[a]{Ranit Das,}
\author[b]{Thorben Finke,}
\author[b]{Marie Hein,}
\author[c,d]{Gregor Kasieczka,}
\author[b]{Michael Kr\"amer,}
\author[b]{Alexander M\"uck} 
\author[a]{and David Shih}

\affiliation[a]{NHETC, Dept.\ of Physics and Astronomy, Rutgers University, Piscataway, NJ 08854, USA}
\affiliation[b]{Institute for Theoretical Particle Physics and Cosmology, RWTH Aachen University, D-52056 Aachen, Germany}
\affiliation[c]{Institut f\"{u}r Experimentalphysik, Universit\"{a}t Hamburg, 22761 Hamburg, Germany}
\affiliation[d]{Center for Data and Computing in Natural Sciences (CDCS), 22607 Hamburg, Germany}

\emailAdd{ranit@physics.rutgers.edu}
\emailAdd{thorben.finke@rwth-aachen.de}
\emailAdd{gregor.kasieczka@uni-hamburg.de}
\emailAdd{marie.hein@rwth-aachen.de}
\emailAdd{mkraemer@physik.rwth-aachen.de}
\emailAdd{mueck@physik.rwth-aachen.de}
\emailAdd{shih@physics.rutgers.edu}

\abstract{
Resonant anomaly detection methods have great potential for enhancing the sensitivity of traditional bump hunt searches.
A key component of these methods is a high quality background template used to produce an anomaly score. Using the LHC Olympics R\&D dataset, we demonstrate that this background template can also be repurposed to directly estimate the background expectation in a simple cut and count setup. In contrast to a traditional bump hunt, no fit to the invariant mass distribution is needed, thereby avoiding the potential problem of background sculpting. Furthermore, direct background estimation allows working with large background rejection rates, where resonant anomaly detection methods typically show their greatest improvement in significance.
}
\begin{document}

\maketitle

\section{Introduction}
\label{sec:Introduction}

The experiments at the Large Hadron Collider (LHC) have advanced the foundations of the Standard Model (SM) by discovering the Higgs boson and providing a wealth of precision measurements. However, the LHC results have not (yet) produced compelling evidence for Beyond the Standard Model (BSM) physics. Numerous searches for BSM models have been conducted and have yielded bounds on the corresponding model parameters, but no evidence that these models are realized in Nature. While it is possible that there is no new physics directly accessible at LHC energies, it is also possible that the BSM physics realized in Nature is not covered by the specific models and signatures under consideration. There is certainly the prospect of exploring more of the model space with specific model building. However, the advent of machine learning (ML) has also greatly enhanced the ability to perform more model-agnostic searches for BSM physics \cite{Kasieczka:2021xcg,Aarrestad:2021oeb, Karagiorgi:2022qnh,Belis:2023mqs, hepmllivingreview}. In particular, resonant anomaly detection techniques \cite{Collins:2018epr, 
Collins:2019jip, 
Nachman:2020lpy, 
Amram:2020ykb, 
Hallin:2021wme, 
Raine:2022hht, 
Hallin:2022eoq,  
Sengupta:2023xqy, 
Buhmann:2023acn, 
Freytsis:2023cjr,
Sengupta:2023vtm, 
Das:2023bcj,
Leigh:2024chm,
sigma,
Andreassen:2020nkr, 
1815227, 
Mastandrea:2022vas, 
Cheng:2024yig, 
Golling:2022nkl, 
Beauchesne:2023vie,
Golling:2023yjq, 
Finke:2023ltw} have the potential to transform simple bump hunts into more powerful multivariate analyses in a model-agnostic manner.

In resonant anomaly detection, one assumes that the signal is localized in some feature $m$, and then the idea is to find ways to estimate the likelihood ratio between data and background (B), 
\begin{equation}
\label{eq:Ropt}
R_\text{optimal}(x)=\frac{p_\text{data}(x)}{p_B(x)}
\end{equation}
in some additional feature space $x$. By the Neyman-Pearson lemma, this is the optimal score for detecting anomalies in the data, and it is completely signal model agnostic. A key step in all resonant anomaly detection methods is to obtain a high-quality model for the distribution of background events (the denominator of Eq.~(\ref{eq:Ropt})). Such a background template could be derived in a data-driven manner from sideband regions
\cite{Collins:2018epr, 
Collins:2019jip, 
Nachman:2020lpy, 
Amram:2020ykb, 
Hallin:2021wme, 
Raine:2022hht, 
Hallin:2022eoq,
Sengupta:2023xqy, 
Buhmann:2023acn, 
Freytsis:2023cjr,
Sengupta:2023vtm, 
Das:2023bcj,
Leigh:2024chm, 
sigma},
or with the help of simulations 
\cite{Andreassen:2020nkr, 
1815227, 
Mastandrea:2022vas, 
Cheng:2024yig, 
Golling:2022nkl, 
Beauchesne:2023vie}.

Applying a cut on the anomaly score can significantly increase the sensitivity to new physics signals by accessing additional features of the events. In order to conclusively detect or reject the presence of new physics in the data, one needs to apply a well-calibrated and robust statistical procedure to the events that survive the cut on the anomaly score. It has generally been assumed that the best approach is to perform a standard bump hunt on the invariant mass distribution of the remaining events -- see for example the existing applications of resonant anomaly detection methods to ATLAS \cite{ATLAS:2021kxv} and CMS \cite{CMS-PAS-EXO-22-026} data.
However, the sculpting of the invariant mass distribution of the background by a cut on the anomaly score can be a potential problem in these approaches.  While ideas have been proposed \cite{Hallin:2022eoq} to overcome the issue of sculpting, these proposals have been limited to specific methods, whereas we present a more general approach.

In this work, we show that resonant anomaly detection methods offer a unique opportunity to find anomalies in an alternative way. Given a suitable background template, the resonance search no longer requires a fit to the invariant mass distribution after data selection. Instead, it can be performed as a very simple counting experiment, where the number of background events in the signal region (defined by a window in $m$) is estimated directly from the background template after applying the anomaly score. 
Direct background estimation was previously studied in the ANODE framework~\cite{Nachman:2020lpy}, but was found to be systematically biased and no proposals were made to deal with the systematic effects. A specific application of direct background estimation, where systematic errors are negligible, has been studied in~\cite{Finke:2022lsu}. Here we explore two different methods for estimating the bias, one simulation-based and the other data-driven, and we show in two examples (CWoLa Hunting \cite{Collins:2018epr,Collins:2019jip} and CATHODE \cite{Hallin:2021wme}) that these methods are sufficient to provide a statistically robust estimate of the number of background events in the signal region.  

With direct background estimation, the problem of potential sculpting of the invariant mass distribution discussed in Ref.~\cite{Hallin:2022eoq} can be completely avoided. Furthermore, direct background estimation can be used for larger background rejection rates where a conventional fit to the invariant mass distribution may be statistically limited. Since this is the region where resonant anomaly detection methods achieve the highest significance improvement, direct background estimation can potentially lead to higher detection sensitivity and better bounds if the background template is good enough, i.e.\ the corresponding systematic error is small enough. Finally, since we introduce a measure of the quality of a given background template, the proposed procedure can be used as a relatively simple way to benchmark weakly supervised resonance search methods using only SM background simulations. 

The proposed analysis is also particularly simple in terms of the statistical procedure used. While there are ideas for more generic or advanced statistical tests in the literature~\cite{DAgnolo:2018cun,dAgnolo:2021aun,Chakravarti:2021svb, Kamenik:2022qxs}, here we simply choose a classifier threshold to select signal-enriched data and perform a standard counting experiment by comparing to our background template expectation. This reduces the statistical power of the analysis, but we show that with the significance improvement of current state-of-the-art anomaly detection methods, our simple method is powerful and robust.

To illustrate our general approach, we will focus on a dijet resonance search using the LHC Olympics R\&D dataset~\cite{Kasieczka:2021xcg, LHCOdataset}. As with previous studies, we stick to the simple case of a few hand-crafted features for an initial proof-of-concept demonstration. However, the setup introduced in this work is generic and not limited to this specific case study. More recently, resonant anomaly detection methods have been extended to larger feature sets and low-level features \cite{Finke:2023ltw, Buhmann:2023acn, Sengupta:2023vtm}. Studying direct background estimation in these more general and model-agnostic settings would be an interesting direction for future study.

The paper is organized as follows. Section~\ref{sec:Setup} introduces the dataset and the weakly supervised method we use. The central idea of performing weakly supervised anomaly detection as a particularly simple cut and count experiment using direct background estimation is described in section~\ref{sec:Sensitivity}. Section~\ref{sec:Results} presents the numerical results of our study, and section~\ref{sec:Conclusion} provides a summary and outlook for future work. In appendix~\ref{app:Architecture} we present the architecture of the classifier and the density estimation used to create a background template in the \Cathode\ framework. A more detailed study of the robustness of our estimate of the systematic uncertainties is presented in appendices~\ref{app:Rsys} and~\ref{app:further_studies}. 
\section{Setup}
\label{sec:Setup}

\subsection{Data set and signal regions}
\label{sec:Data}

As in previous studies of resonant anomaly detection, we use the R\&D data set of the LHC Olympics \cite{Kasieczka:2021xcg, LHCOdataset}. The data set consists of $10^6$ QCD dijet events with a leading jet transverse momentum $p_\mathrm{T}> 1.2~\text{TeV}$. As in Ref.~\cite{Hallin:2021wme}, in addition to the dijet mass, $m_{JJ}$, we use the invariant mass $m_{J_1}$ of the leading jet, the mass difference $\Delta m = m_{J_2} - m_{J_1}$ between the leading and subleading jets, and the ratio of the 1-subjettiness and 2-subjettiness of the two leading jets, $\tau_{21}^{J_1}$ and  $\tau_{21}^{J_2}$ with $\tau_{ij} \equiv \tau_i/\tau_j$, as a baseline feature set. We also study a variation of this feature set where we add the angular distance between the two jets, $\Delta R = \sqrt{(\phi_{J_2} - \phi_{J_1})^2 + (\eta_{j_2} - \eta_{J_1})^2}$, a feature correlated with the dijet mass. This feature has been used previously in studies of correlations of classification features with the dijet mass, e.g.\ in Refs.~\cite{Raine:2022hht, Hallin:2022eoq}. 
The events of the LHC Olympics R\&D data set were generated using Pythia~8.219 \cite{Sjostrand:2014zea} and Delphes~3.4.1 \cite{deFavereau:2013fsa} with default settings. To assess the robustness of our method against an imperfect Monte Carlo simulation of the background data, we also use a generation of $10^6$ QCD dijet events using Herwig++~\cite{Bahr:2008pv} and Delphes with default settings and cuts as above, as published in the LHC Olympics as Black Box 2 \cite{Kasieczka:2021xcg, LHCOblackboxes}. 

The new physics signal of the R\&D data set is a $Z'$ resonance with mass $m_{Z'}=3.5~\text{TeV}$ which decays into two bosons $X$ and $Y$ with masses $m_{X}=500~\text{GeV}$ and $m_{Y}=100~\text{GeV}$, respectively. The $X$ and $Y$ bosons each decay promptly to pairs of quarks. We use this signal as a benchmark for the anomaly detection methods under investigation. 

To perform an idealized analysis (employing a perfect background template) with the same statistics we need to double the data set size as discussed in Section~\ref{sec:Idealized}. For this purpose, we have generated an additional data set with $2\cdot 10^6$ background and $10^5$ signal events with the settings of the LHC Olympics R\&D data set. 

For the resonance search, we center overlapping signal regions of width 400~GeV at the dijet invariant masses $m_n=(3.5-0.1\cdot (5-n))~\text{TeV}$ with $n=1,\ldots, 9$. Towards higher invariant masses one runs out of statistics, and towards lower invariant masses we are limited by the generation cut $p_\mathrm{T}> 1.2~\text{TeV}$ for the leading jet. We perform a sliding window analysis and expect to identify an anomaly only in signal regions containing the benchmark signal. In our setup, the center of the signal region $n=5$ coincides with the resonance mass of our benchmark signal. 

We use two settings for our analyses: A background-only setting, where no signal is injected, to test the compatibility of our setup with the null hypothesis and a benchmark point with a signal injection of 1000 events ($N_S/\sqrt{N_B}\approx 2.2$ in window 5) to test the sensitivity. 

K-fold cross-validation is used to obtain independent training and test sets while still utilizing the full statistics of the data. We use $k=5$ with four folds forming our training and validation sets and one fold forming the test set. For each window, five classifiers - one per fold - are then trained so that classifier scores are obtained for all events in the data set.  

\subsection{Weakly supervised setup}
\label{sec:weakly_supervised}

In the weakly supervised setup, a supervised classifier is trained on mixed instead of pure data sets. For two data sets with the data distributions $p_i(x)=f_i\, p_S(x)+(1-f_i)\, p_B(x)$, where $f_i$ is the signal fraction and $p_{S/B}(x)$ are the signal/background distributions, an optimal classifier is the likelihood ratio~\cite{Neyman:1933wgr}
\begin{equation}
    R(x) = \frac{p_1(x)}{p_2(x)} = \frac{f_1\, p_S(x)/p_B(x)+(1-f_1)}{f_2\, p_S(x)/p_B(x)+(1-f_2)}\,.
\end{equation}
This classification task is equivalent to supervised classification since there is a monotonic relation of $R$ to the optimal classifier in the supervised case, $p_S(x)/p_B(x)$. In these weakly supervised methods, we therefore attempt to obtain such mixed data sets. In the methods we use, this is done by using the signal region data ($p_1(x)=p_\text{data}(x)$), which might contain signal and background events, as a signal-enriched dataset, and constructing a background template ($p_2(x)\approx p_B(x)$) from the sidebands, where only background events should be present. If there is no signal, the classifier can only guess at random if the probability densities of the two samples are identical. If there is signal, however, the classifier will learn the signal versus background classification through this proxy task, as this is the only way to distinguish the two data sets. In practice, if the background template is not ideal, there may be slight differences between the distributions of the background events of the two datasets that need to be accounted for in later steps of the analysis (see Section~\ref{sec:Sensitivity}). 

The weakly supervised setup requires powerful classification algorithms capable of identifying signal events for small signal fractions. For high-level features, boosted decision trees (BDTs) have recently been identified as such a robust classification architecture~\cite{Finke:2023ltw, Freytsis:2023cjr}. We use the BDT-based classifier of Ref.~\cite{Finke:2023ltw}, which uses an ensemble of 50 individual BDTs for classification. Details of the architecture and training procedure can be found in Appendix~\ref{app:bdt}. 

In the following subsections, we briefly discuss the methods to obtain a background template which are used in our case study in Section~\ref{sec:Results} for the generic setup introduced in Section~\ref{sec:Sensitivity}. We employ the idealized anomaly detector, CWoLa and \Cathode. However, we want to stress that the generic setup can be used with any method to obtain a background template.

\subsubsection{Idealized anomaly detector}
\label{sec:Idealized}

For the Idealized Anomaly Detector (IAD), we assume that we have a perfect background template whose feature space probability density exactly matches the probability density of the background in the signal region. With simulated data, unlike real data, we can simply use Monte Carlo generated background events in the signal region.  

Since the background template also consists of simulated data in this setup, we need to double the size of the data set to obtain results with the same statistics. Hence, for our IAD studies, we use $2\cdot 10^6$ events we have generated ourselves (see Section~\ref{sec:Setup}). We use the first $10^6$ background events to replace the LHCO R\&D dataset and the rest as a background template of the same size.

\subsubsection{CWoLa}
\label{sec:CWoLa}

For CWoLa, the background template consists of short sidebands of width 0.2~TeV on either side of each signal region. If the features used were uncorrelated with the dijet mass $m_{JJ}$ and the sidebands contained no signal events, the CWoLa setup would be identical to the idealized anomaly detector in Sec.~\ref{sec:Idealized}. Correlations degrade detection performance, and the associated systematic uncertainties must be taken into account to avoid false anomaly detections. We use only the short sidebands as a background template to limit the effect of potential correlations. 

\subsubsection{\Cathode}
\label{sec:Cathode}

For the \Cathode{} method, it is not necessary to assume that the additional classification features, e.g.\ $x = (m_{J_1},\Delta m,\tau_{21}^{J_1},\tau_{21}^{J_2},\Delta R)$, and the invariant mass $m_{JJ}$ are uncorrelated. Instead, we make the weaker assumption that the conditional probability distribution $p(x|m_{JJ})$ is smooth with respect to $m_{JJ}$. The probability distribution $p(x|m_{JJ})$ is learned on the sidebands using density estimation conditioned on $m_{JJ}$.  The density estimator is then interpolated into the signal region and used as a generative model to sample background events for the background template. This allows us to generate a large sample of events to avoid statistical limitations. This is commonly referred to as oversampling. The probability distribution $p(m_{JJ})$ for the invariant mass in the signal region is estimated by kernel density estimation as in Ref.~\cite{Hallin:2021wme}. In contrast to the normalizing flow architecture in Ref.~\cite{Hallin:2021wme}, in this work, we employ a more expressive and faster-to-train density estimator, Conditional Flow Matching (CFM) \cite{sigma, lipman2023flowmatchinggenerativemodeling}. Details about CFM and the training procedure are described in Appendix~\ref{app:flow}. 

For \Cathode, the sidebands consist of all data except the data in the signal region of interest. A fixed oversampling factor of four is used for classification. 
\section{Direct background estimation for resonant anomaly detection}
\label{sec:Sensitivity}

After setting up a background template with $N_\text{BT}$ events, as discussed in section~\ref{sec:Setup} for the different methods, the resonance search can be performed in a straightforward way for each signal region with $N_\text{SR}$ events. We choose a background efficiency $\epsilon_B$ and use a working point $R_c$ of the weakly supervised classifier such that $\epsilon_B N_\text{BT}$ events of the background template are incorrectly classified as signal region events.\footnote{Since we employ $k$-fold cross-validation, the working points for the different classifiers are chosen individually so that each results in an efficiency of $\epsilon_{B}$ on the corresponding $k$-fold of the background.} For the given cut on the anomaly score, some number of events $N_{\rm obs}$ with $R(x)>R_c$ survive in the data. In direct background estimation, we aim to derive a background-only expectation $N_\text{exp}$ for $N_{\rm obs}$ from the background template. If the background template were perfect, 
$\epsilon_B N_\text{SR}$ would be the background estimate for the data. However, in a realistic analysis, the background template will always differ from the true one in some systematic ways; this will bias the background expectation accordingly:
\begin{equation}\label{eq:Nexp}
N_\text{exp}=\epsilon_B N_\text{SR}\left(1+\delta_\text{sys}(\epsilon_B) \right) \, .
\end{equation}
Hence, for direct background estimation it is essential to estimate the systematic shift $\delta_\text{sys}$ and its systematic uncertainty $\sigma_\text{sys}$ for a given $\epsilon_B$ in a well controlled way. In the following subsections, we will present two different methods for estimating $\delta_{\rm sys}$: one simulation-based and the other data-driven. 

Note that $\delta_\text{sys}$ (and our proposed methods for estimating it) also provides a direct measure of the quality of the background template. Improving the background template reduces $\delta_\text{sys}$ and the IAD case with $\delta_\text{sys}=0$ is approached. Obviously, the goal is to find a background template with $\delta_\text{sys}$ as small as possible.

\subsection{Method 1: MC-based estimate of $\delta_\text{sys}$}
\label{sec:systematic_shift_and_error}

For an actual analysis, one needs to establish a robust procedure for estimating $\delta_\text{sys}$ and the corresponding systematic uncertainty $\sigma_\text{sys}$. 
We propose to use $\delta_\text{sys}=\delta_\text{sys}^\text{MC}$, where $\delta_\text{sys}^\text{MC}$ is determined from signal-free Monte Carlo data. To investigate the robustness of our procedure against an imperfect Monte Carlo simulation of the background data, we determine $\delta_\text{sys}^\text{MC}$ from one million QCD dijet events simulated with the Herwig event generator instead of Pythia. Even for a signal-free (MC) data set, due to limited statistics, we can only estimate $\delta_\text{sys}$. We use the following procedure: For each signal region $n=1,\ldots 9$ in our sliding window search, we calculate the ratio 
\begin{equation}\label{eq:R}
    \delta_{\text{sys},n}=\frac{N_{\text{obs},n}-\epsilon_B\, N_{\mathrm{SR},n}}{\epsilon_B\, N_{\mathrm{SR},n}}
\end{equation}
by averaging the results for 10 classifiers. Since the MC is signal free, $N_\text{obs}=N_\text{exp}$ holds up to statistical fluctuations and \ref{eq:Nexp} and \ref{eq:R} are equivalent. For \Cathode, each classifier is trained using a background template generated by an independent density estimator. Our estimate of $\delta_\text{sys}$ is defined as the average of $\delta_{\text{sys},n}$ over all signal windows, i.e.
\begin{equation}\label{eq:delta sys}
    \delta_{\text{sys}}=\frac{1}{9} \sum_{n=1}^9 \delta_{\text{sys},n} \, .
\end{equation}

In contrast to a real analysis, we can use this procedure to find $\delta_\text{sys}^\text{data}$ on our (labeled Pythia) data set by investigating only background data. We consider $\delta_\text{sys}^\text{data}$ as a benchmark for the estimates of $\delta_\text{sys}$ that are available in a real analysis and are discussed in the following.

For simplicity, we also use $\delta_\text{sys}^\text{MC}$ as the relative systematic error, i.e.\ $\sigma_\text{sys}=\delta_\text{sys}=\delta_\text{sys}^\text{MC}$. The observed fluctuations of $\delta_{\text{sys},n}$ (see \ref{eq:R}) for our Monte Carlo data, which are discussed further in Appendix~\ref{app:Rsys}, are largely covered by the expected statistical fluctuations. Hence, for a given data set, a systematic error as large as $\delta_\text{sys}$ seems to be conservative. However, we also use $\sigma_\text{sys}=\delta_\text{sys}$ to account for the expected differences between Monte Carlo and data. For increasing $\delta_\text{sys}^\text{MC}$, we expect the difference to the true $\delta_\text{sys}$ to also increase. 

\subsection{Method 2: Data-driven estimate of $\delta_\text{sys}$ from sidebands}
\label{sec:deltasysSB}

Using data-driven methods to estimate $\delta_\text{sys}$ and $\sigma_\text{sys}$, rather than relying solely on Monte Carlo studies, is an additional way to validate and improve the error estimate. In the  \Cathode{} framework, we create an alternative background template by generating sideband data from our density estimator, i.e.\ in contrast to our standard approach, we do not interpolate into the signal region. By performing a classification of this background template against the sideband data, which might also contain signal, we define $\delta_\text{sys}^\text{SB}$ as described above. 

As usual, we use the full sideband data to train the generative network in order to improve the training and to reduce the influence of a localised signal on the learning of the background data distribution. The classifier is trained using the same data set sizes as for the signal region analysis. We take a randomized selection of the full sideband data for each of the 10 classifier runs and oversample the corresponding background template by a factor of four as usual. 

The shift $\delta_\text{sys}^\text{SB}$ provides an estimate of the mismodeling of the background template by the generative network directly on data, and hence of $\delta_\text{sys}$, as long as the interpolation between the sideband and signal regions works well. The mismodeling error is dominant for small $\epsilon_B$, where the tails of the probability distribution need to be estimated. The interpolation error between the sideband and signal regions can be estimated by MC. We simply add these two sources for $\delta_\text{sys}$ in quadrature and suggest using  $\delta^{\text{MC}\oplus\text{SB}}_\text{sys}=\sqrt{\left(\delta^{\text{MC}}_\text{sys}\right)^2+\left(\delta^{\text{SB}}_\text{sys}\right)^2}$ as an additional robust estimate of the systematic shift. 

\subsection{Statistical treatment of the cut-and-count experiment}

Since our setup is a simple cut and count analysis, also the statistical interpretation is simple when $N_\text{obs}$ instead of $N_\text{exp}$ events are observed in a given signal region. We use the Asimov estimate~\cite{Cowan:2010js} for the significance
\begin{equation}\label{eq:significance}
    \mathcal{S} = 
    \left[
    2\left(
    N_\text{obs}\ln\left[\frac{N_\text{obs}(N_\text{exp}^{-1}+\sigma_\text{exp}^2)}{1+N_\text{obs}\sigma_\text{exp}^2}\right]
    -\frac{1}{\sigma_\text{exp}^2}\ln\left[
        \frac{1+N_\text{obs}\sigma_\text{exp}^2}{1+N_\text{exp}\sigma_\text{exp}^2}
    \right]
    \right)    
    \right]^{1/2}\,,
\end{equation}
where $\sigma_\text{exp}$ is the relative error on the background estimate $N_\text{exp}$. If $N_\text{obs}-N_\text{exp}\ll N_\text{exp}$, which we expect to be the case for smaller background rejection $1/\epsilon_B$, this formula reduces to the well-known Gaussian limit
\begin{equation}\label{eq:significance_Gaussian_limit}
    \mathcal{S}_G=\left[ \frac{(N_\text{obs}-N_\text{exp})^2}{(N_\text{exp}+N_\text{exp}^2\sigma_\text{exp}^2)} \right]^{1/2}.
\end{equation}

The error $\sigma_\text{exp}$ in Eqn.~(\ref{eq:significance}) consists of the systematic error $\sigma_\text{sys}$ on the determination of $\delta_\text{sys}$ and an additional statistical error $\sigma_\text{exp,stat}$, since the determination of $\epsilon_B$ from the background template is itself statistics limited. We also treat both $\sigma_\text{exp,stat}$ and $\sigma_\text{sys}$ as relative errors and use
\begin{equation}
    \sigma_\text{exp}=\sqrt{\sigma_\text{exp,stat}^2+\sigma_\text{sys}^2}
\end{equation}with $\sigma_\text{exp,stat}=1/\sqrt{\epsilon_B\, N_\text{BT}}$. For CWoLa, where $N_\text{BT}\approx N_\text{SR}$, $\sigma_\text{exp,stat}$ is about the same size as the statistical error on $N_\text{exp}$. For \Cathode, oversampling can reduce this additional source of error to a negligible amount ($\sigma_\text{exp,stat}\approx 0$).

It is always interesting to compare the systematic error $\sigma_\text{sys}$ with the total statistical error due to the limited data set size. According to the discussion in the previous paragraph, the total relative statistical error is given by 
\begin{equation}\label{eq:sigma stat}
\sigma_\text{stat}=\sqrt{(N_\text{exp})^{-1}+(\epsilon_B\, N_\text{BT})^{-1}}
\end{equation}
as we assume Poisson distributed event numbers. We will compare $\sigma_\text{stat}$ to $\sigma_\text{sys}$ in Section~\ref{sec:Results}.

\section{Results}
\label{sec:Results}

In this section, we will demonstrate the efficacy of the general methods for estimating $\delta_{\rm sys}$ described above, using the Idealized Anomaly Detector, CWoLa Hunting and \Cathode{} as three representative examples of weakly-supervised anomaly detection. 

\subsection{Idealized Anomaly Detector}
\label{subsec:IAD}

For the idealized anomaly detector, there are no mismodeling issues in the background template; we are only limited by the finite statistics of the data and the background template. Hence, $\delta_\text{sys}=\sigma_\text{sys}=0$ is the consistent choice. This expectation is confirmed by our calculations for $\delta_\text{sys}^\text{data}$, see Table~\ref{tab: Rsys all} in Appendix~\ref{app:Rsys}. The small, nonzero values observed for $\delta_\text{sys}$ are only due to the inability of the classifier to perfectly learn the likelihood ratio due to finite training statistics and model capacity. 

\begin{figure}[t]
    \centering
    \subfloat[IAD: $S/B=0\%$]{\includegraphics[width=0.49\textwidth]{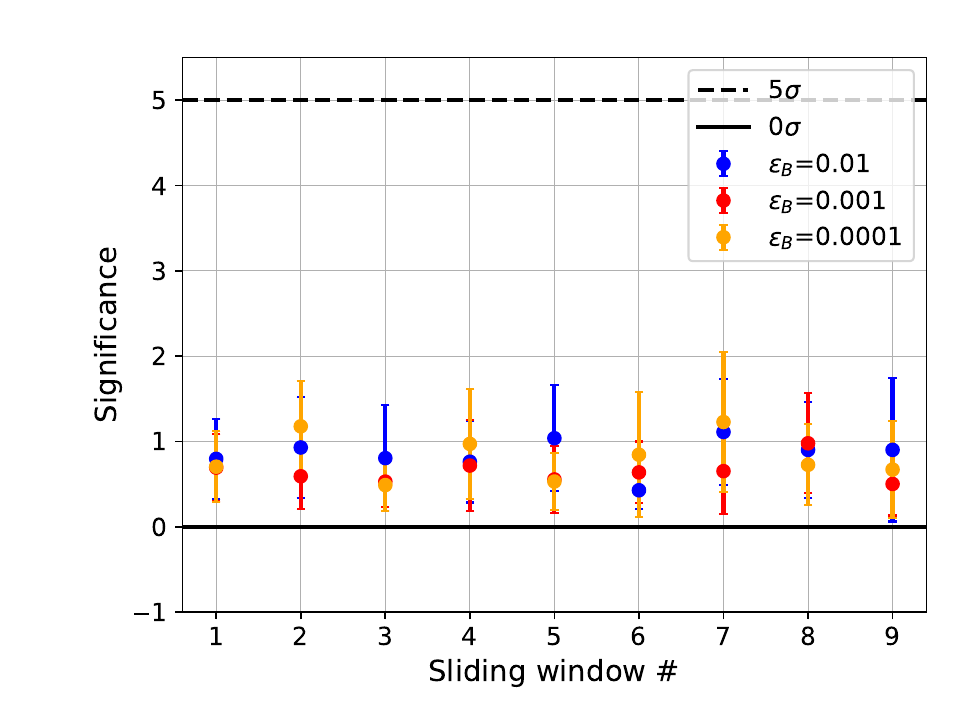}}
    \subfloat[IAD $\Delta R$: $S/B=0\%$]{\includegraphics[width=0.49\textwidth]{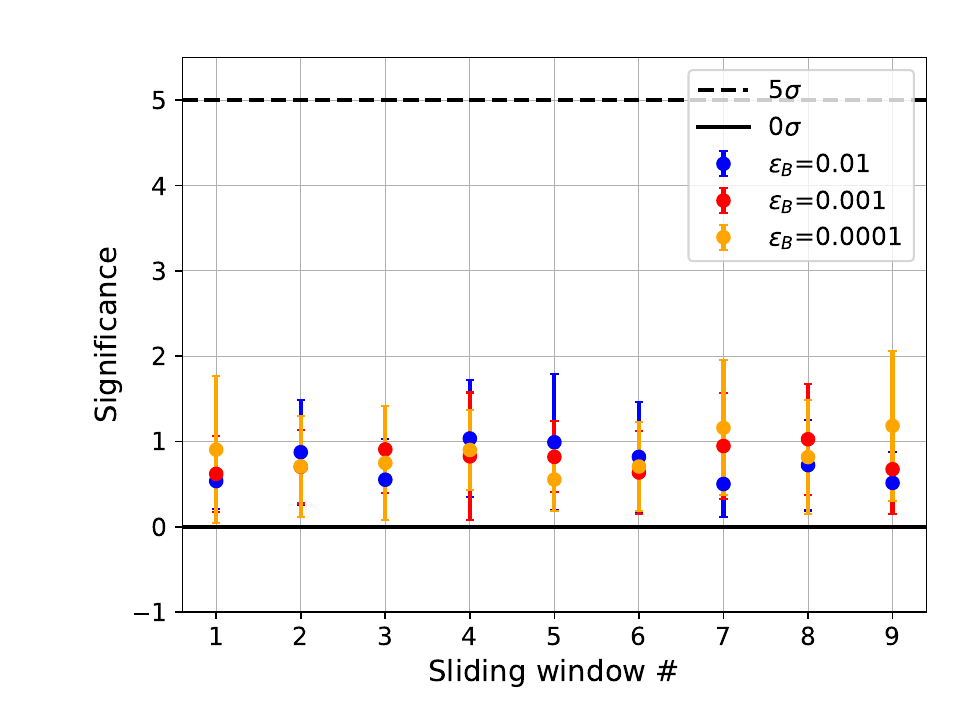}}
    \\
    \subfloat[IAD: $S/B=0.64\%$]{\includegraphics[width=0.49\textwidth]{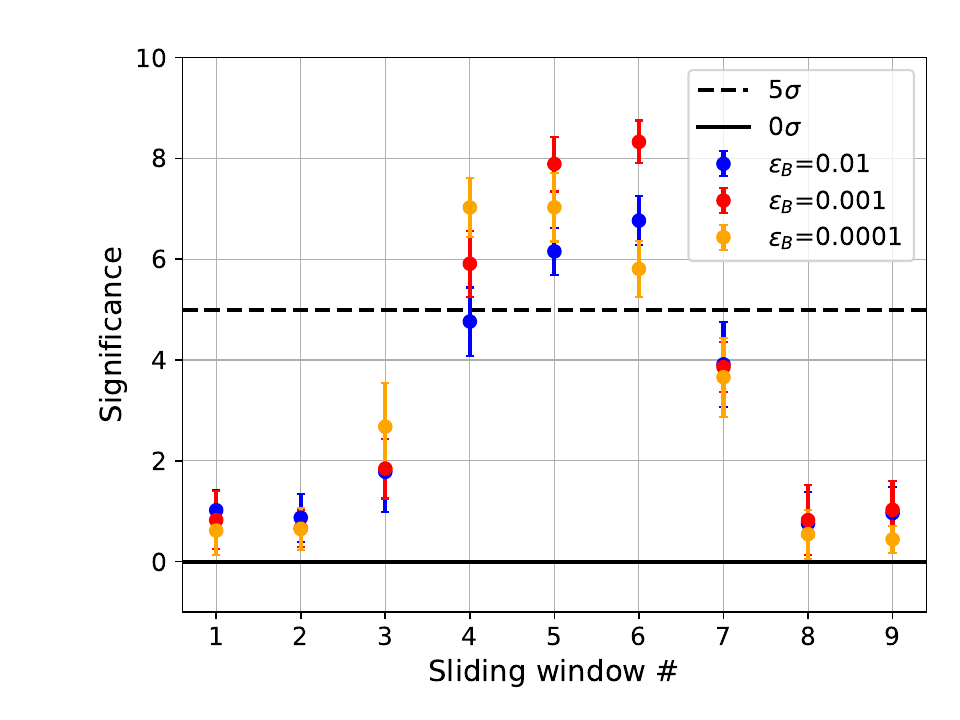}}
    \subfloat[IAD $\Delta R$: $S/B=0.64\%$]{\includegraphics[width=0.49\textwidth]{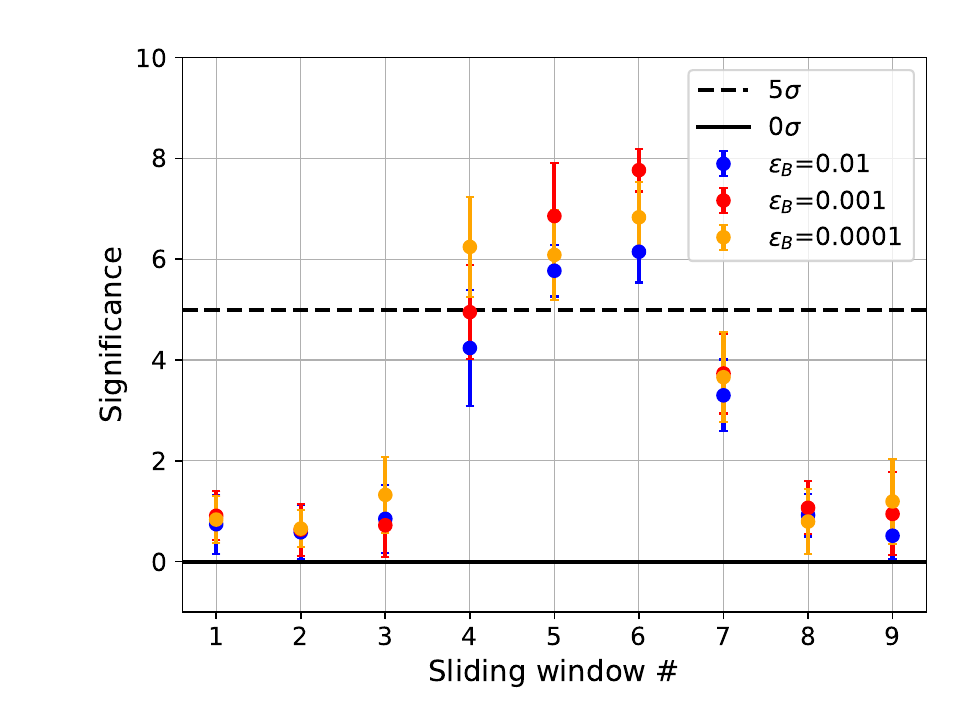}}
    \caption{Significance ${\cal S}$, eqn.~(\ref{eq:significance}), for the different signal regions for the IAD without (top) and with signal injection (bottom) using the baseline dataset (left) and the dataset with $\Delta R$ (right) with $\delta_\text{sys}=\sigma_\text{sys}=0$. The error bars indicate the variance of the significance based on 10 classifier runs.}
    \label{fig:significances IAD}
\end{figure}

When no signal is injected, the observed significances for the signal windows simply show the expected distribution due to statistical fluctuations as shown in Figure~\ref{fig:significances IAD}, panels (a) and (b). An injected signal is discovered with a large significance at all working points, see  Figure~\ref{fig:significances IAD}, panels~(c) and (d). For example we observe an $8\sigma$ discovery for $\epsilon_B=10^{-3}$ using the baseline feature set without $\Delta R$, see  Figure~\ref{fig:significances IAD}, panel~(c). A further discussion of the observed and expected significance, including comparisons with previous studies, can be found in Appendix \ref{sec:significances IAD}.

\subsection{CWoLa Hunting}
\label{subsec:CWoLa}

For CWoLa (as well as for \Cathode{} as discussed in Section~\ref{subsec:Cathode}) on the baseline feature set, using $\delta_\text{sys}=0$ would lead to false discoveries, especially for large $\epsilon_B$ where the statistical error is quite small. On the other hand, for small $\epsilon_B$, statistics becomes a limiting factor in estimating $\delta_\text{sys}$. The results for $\delta_\text{sys}^\text{MC}$ using Herwig MC as well as the reference value $\delta_\text{sys}^\text{data}$ obtained using (Pythia) data are shown in Figure~\ref{fig:Rsys cwola}, see also Table~\ref{tab: Rsys all} in Appendix~\ref{app:Rsys}. We find moderate values of $\delta_\text{sys} \approx 0.2$ or less. The MC estimate $\delta_\text{sys}^\text{MC}$ slightly underestimates $\delta_\text{sys}^\text{data}$ for medium and small $\epsilon_B$. However, the difference is covered by the combination of using $\sigma_\text{sys}=\delta_\text{sys}$ and the increasing statistical error for small $\epsilon_B$ (see the lower panel).

\begin{figure}[t]
    \centering
    \subfloat[CWoLa Baseline]{\includegraphics[width=0.49\textwidth]{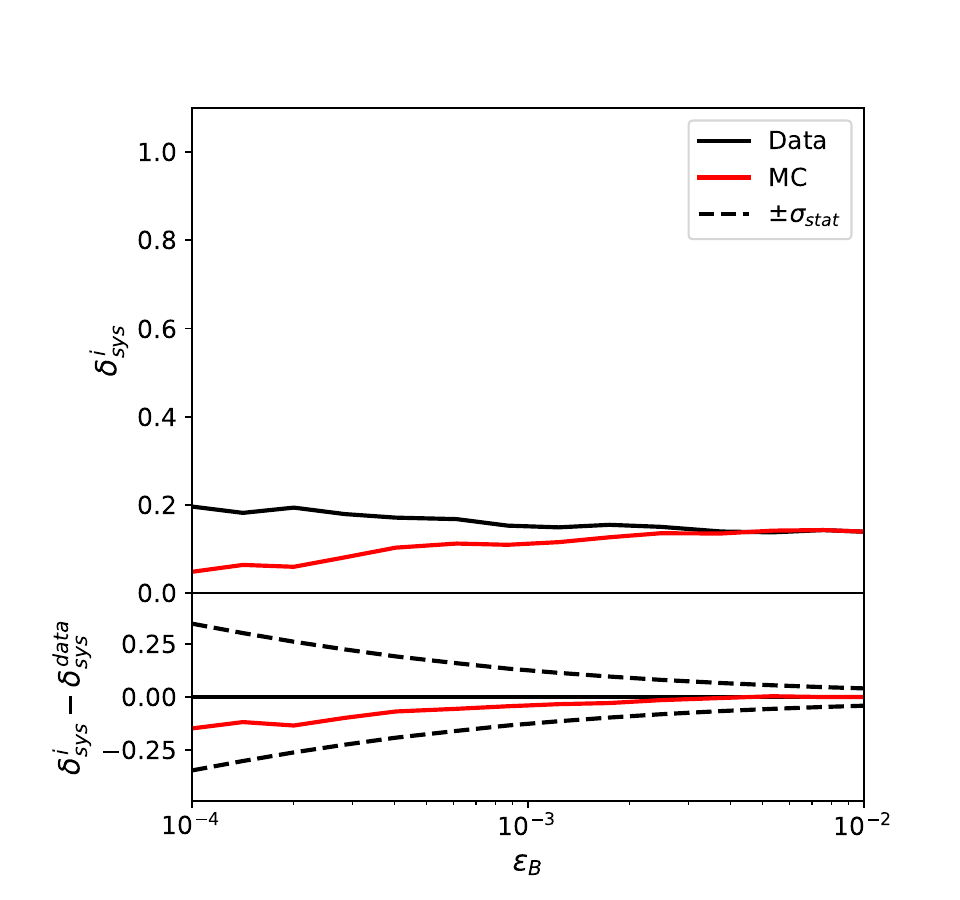}}  
    \subfloat[CWoLa $\Delta R$]{\includegraphics[width=0.49\textwidth]{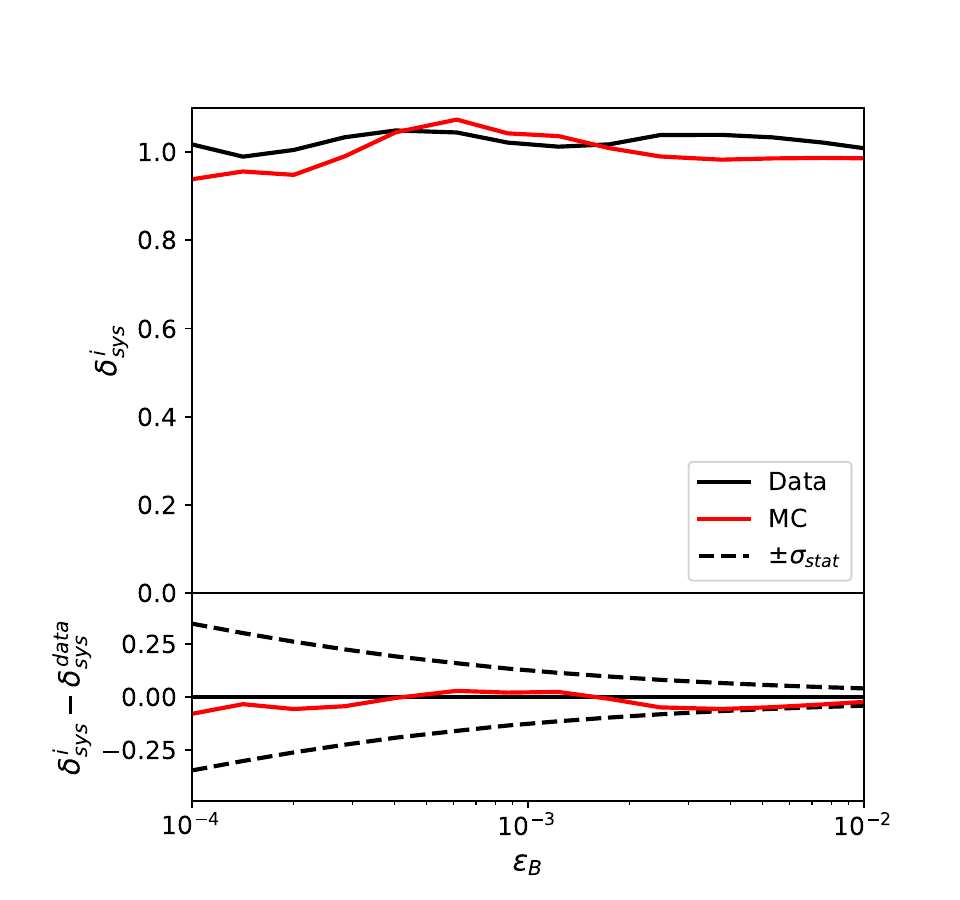}}\\
    \caption{Relative systematic shift $\delta_\text{sys}$ for CWoLa, as a function of $\varepsilon_B$ for the baseline dataset (left) and the dataset with $\Delta R$ (right). $\delta_\text{sys}$ has been estimated from an analysis without signal on Pythia data (Data) and Herwig MC (MC). The total statistical error $\sigma_\text{stat}$ defined in eqn.~(\ref{eq:sigma stat}) is also shown in the lower panel to guide the eye as to how relevant the observed deviations between $\delta_\text{sys}^\text{MC}$ and $\delta_\text{sys}^\text{data}$ are. Note that $\sigma_\text{stat}$ is not an error on $\delta_\text{sys}$. The results are based on 10 classifier runs in each signal region.}
    \label{fig:Rsys cwola}
\end{figure}

If we add $\Delta R$ to the feature set, which is known to correlate with the dijet invariant mass, the picture changes drastically, see the right panel in Figure~\ref{fig:Rsys cwola}. We find a large value of $\delta_\text{sys}\approx 1$, showing that the CWoLa method is hard to control in this case. Note that studying $\delta_\text{sys}$ alone tells us that CWoLa cannot be used with $\Delta R$ as an additional feature. It is not necessary to study a specific signal model to reach this conclusion. However, the MC estimate $\delta_\text{sys}^\text{MC}$ is in good agreement with the reference value $\delta_\text{sys}^\text{data}$ both within the statistical error of the analysis $\sigma_\text{sys}$ and within the dramatically enlarged systematic error, which follows from our error estimate. Therefore, we do not expect to see any significant deviation from the background only hypothesis in our analysis without signal. 

\begin{figure}[t]
    \centering
    \subfloat[CWoLa: $S/B=0\%$]{\includegraphics[width=0.49\textwidth]{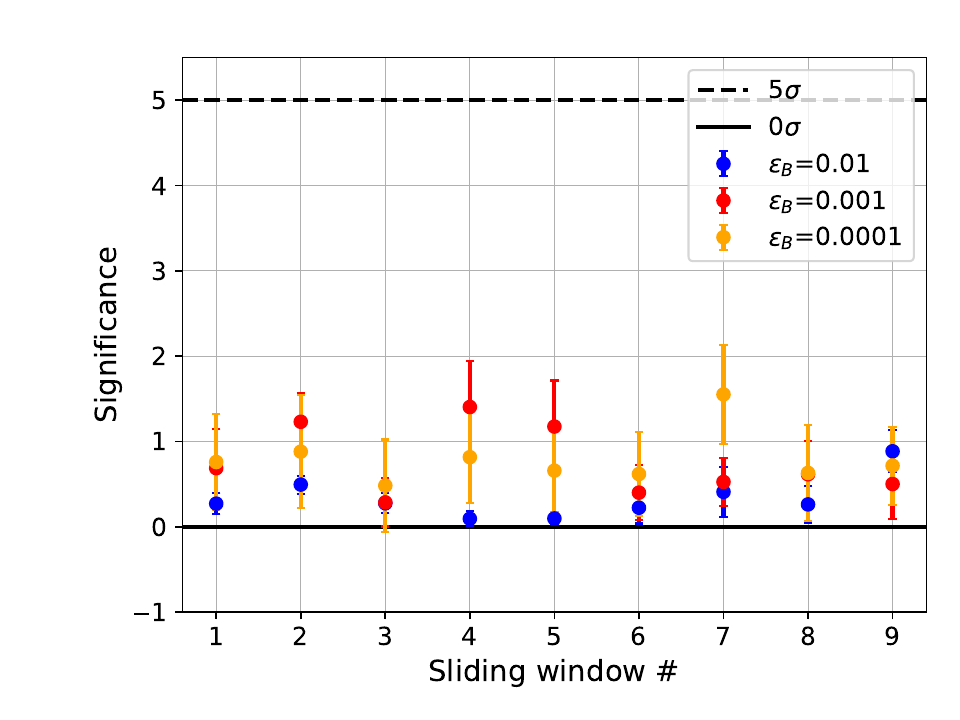}}  
    \subfloat[CWoLa $\Delta R$: $S/B=0\%$]{\includegraphics[width=0.49\textwidth]{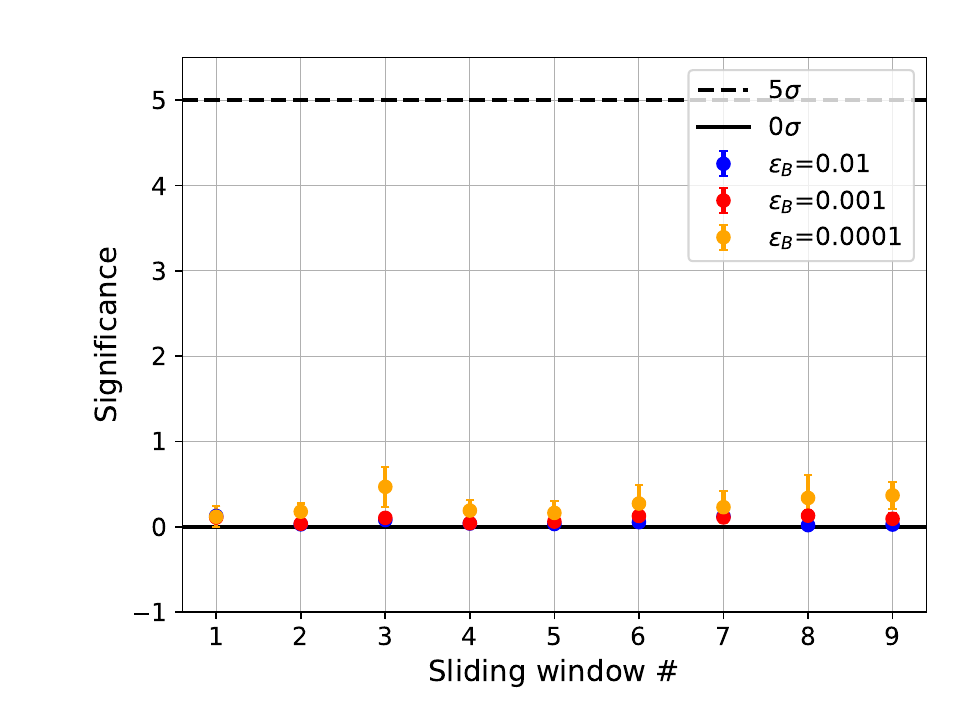}}\\
    \subfloat[CWoLa: $S/B=0.64\%$]{\includegraphics[width=0.49\textwidth]{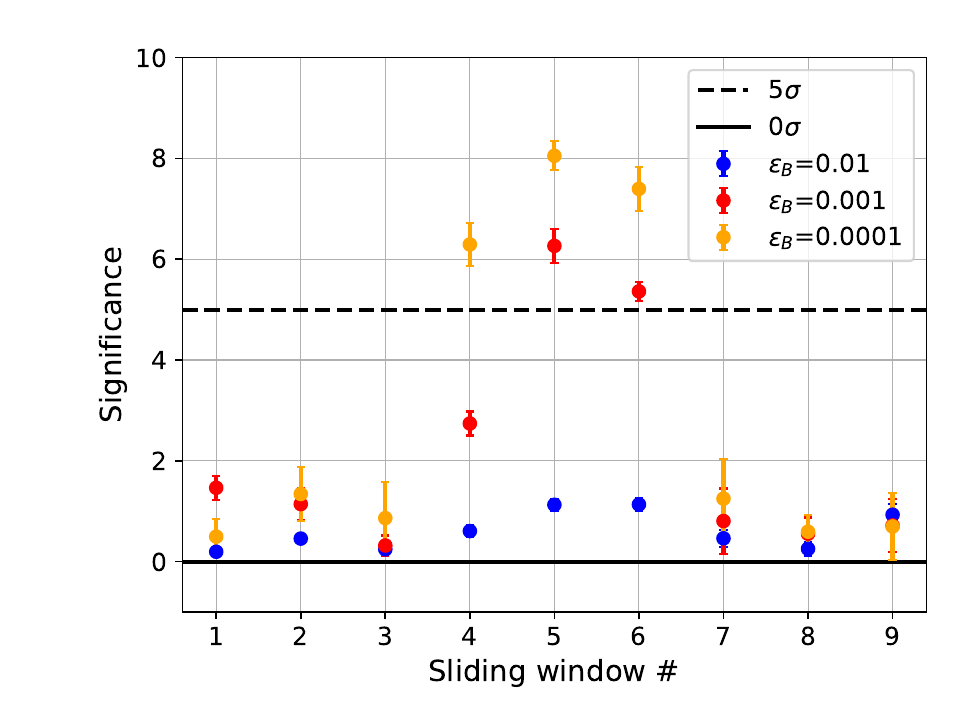}}
    \subfloat[CWoLa $\Delta R$: $S/B=0.64\%$]{\includegraphics[width=0.49\textwidth]{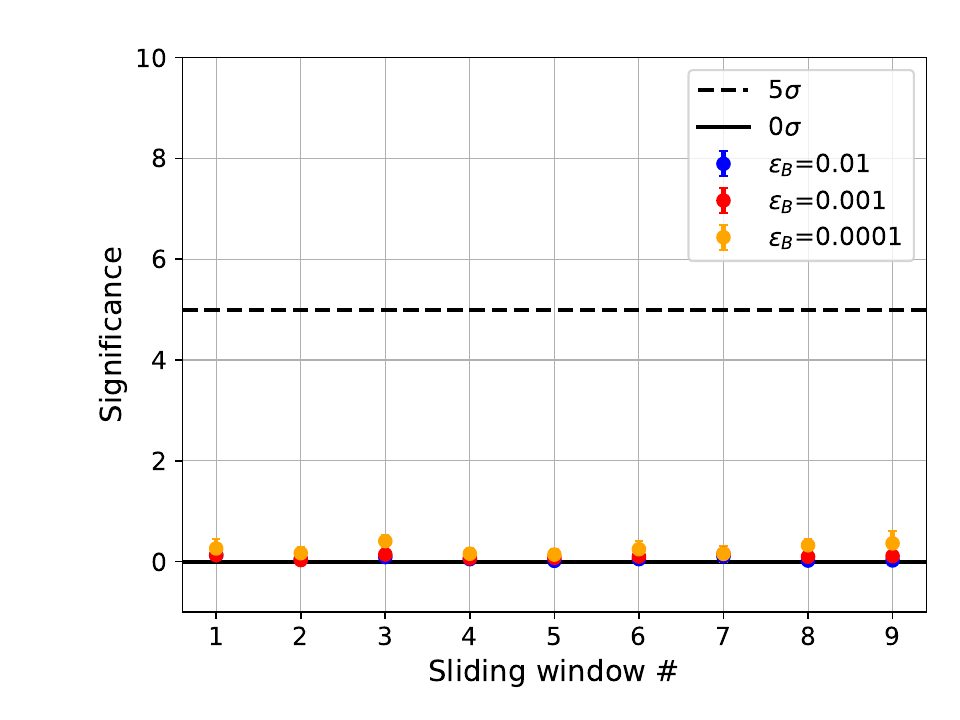}}
    \caption{Significance ${\cal S}$, eqn.~(\ref{eq:significance}), for the different signal regions for CWoLa without (top) and with signal injection (bottom) using the baseline dataset (left) and the dataset with $\Delta R$ (right). The error bars indicate the variance of the significance based on 10 classifier runs.}
    \label{fig:significances cwola}
\end{figure}
This is confirmed by Figure~\ref{fig:significances cwola}, panels (a) and (b), where we show the results of such an analysis for the CWoLa approach using the baseline feature set and the feature set including $\Delta R$. None of the potential signal windows show a significant deviation from the background only hypothesis. Also for CWoLa with $\Delta R$, where the correlation of a classification feature with the dijet mass leads to a breakdown in performance, the analysis itself is robust and no false discovery is observed. 

In Figure~\ref{fig:significances cwola}, panels (c) and (d), we show the results of the cut and count analysis for the standard signal injection. For the baseline feature set, we are able to achieve an improved significance of more than $5\sigma$ for two working points ($\epsilon_B=10^{-3}$ and $10^{-4}$) for CWoLa. For $\epsilon_B=10^{-2}$ even the moderate systematic error $\sigma_\text{sys}=0.14$ is much larger than the statistical error. Therefore, our setup will not be as sensitive at $\epsilon_B=10^{-2}$. For $\epsilon_B=10^{-3}$ the estimated systematic error is of the same size as the statistical error, while for $\epsilon_B=10^{-4}$ the systematic error is almost negligible. Hence, in contrast to the IAD analysis, the working point $\epsilon_B=10^{-4}$ leads to the largest significance.

As expected from the determination of $\delta_\text{sys}$, we observe a complete failure of CWoLa in the analysis with $\Delta R$. We do not cross the $5\,\sigma$ boundary with any of the three thresholds, and in contrast to panel (c), where we still observe the bump at $\epsilon_B=10^{-2}$, we see no such structure here. The classifier largely classifies events according to $\Delta R$ and therefore does not perform the signal versus background classification necessary to obtain a significant deviation here. However, as shown in our previous discussion, we do not see any false positives resulting from this breakdown of the method.

\subsection{\Cathode}
\label{subsec:Cathode}

\begin{figure}[t]
    \centering
    \subfloat[CATHODE Baseline]{\includegraphics[width=0.49\textwidth]{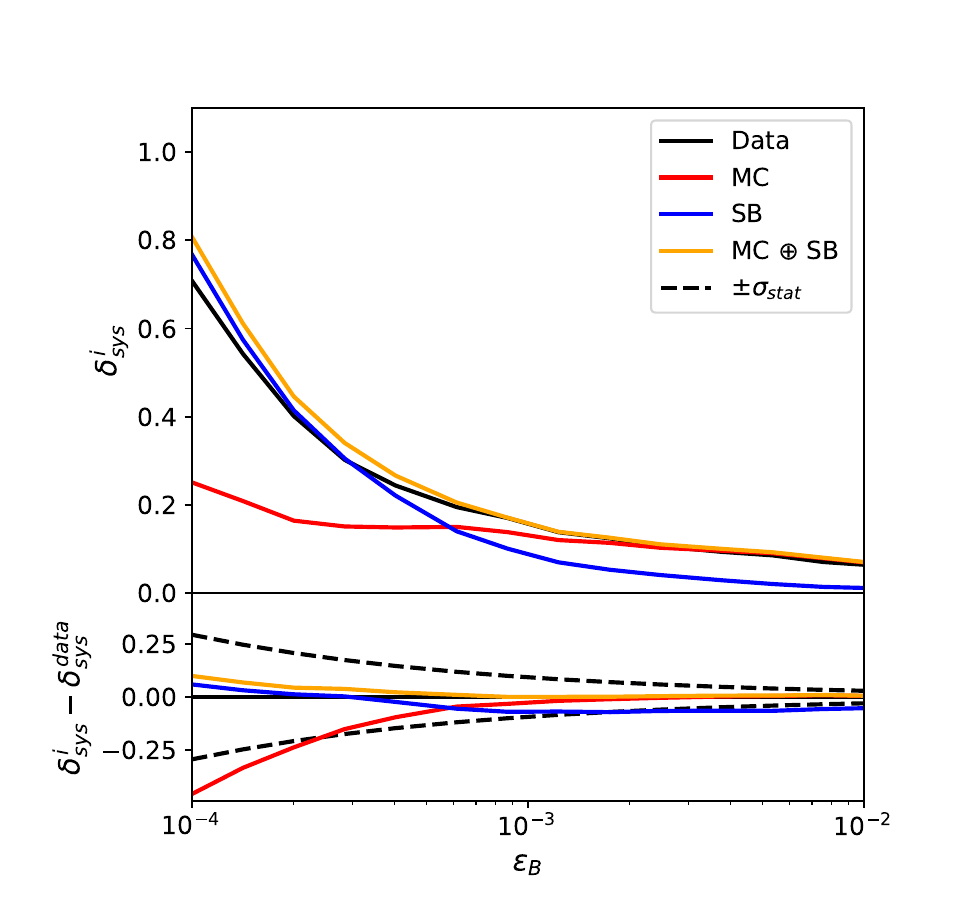}}  
    \subfloat[CATHODE $\Delta R$]{\includegraphics[width=0.49\textwidth]{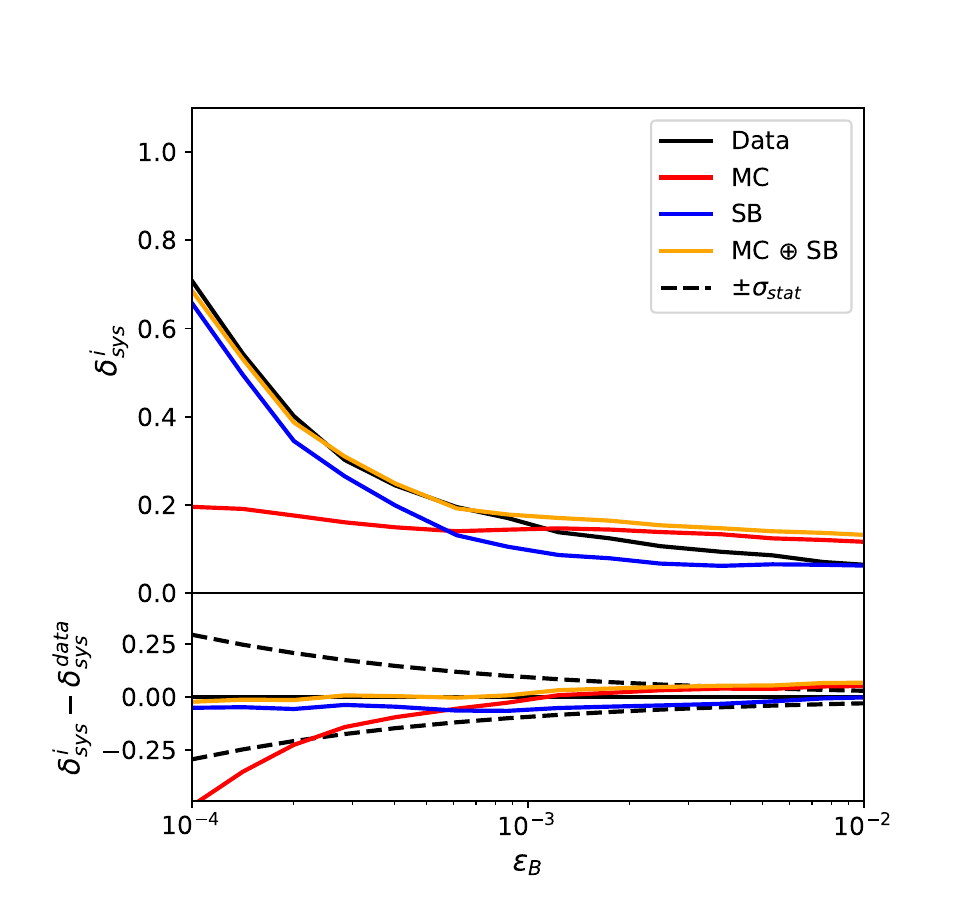}}\\
    \caption{Relative systematic shift $\delta_\text{sys}$ for \Cathode, as a function of $\varepsilon_B$ for the baseline dataset (left) and the dataset with $\Delta R$ (right). $\delta_\text{sys}$ has been estimated from an analysis without signal on Pythia data ($\delta_\text{sys}^\text{data}$), Herwig MC ($\delta_\text{sys}^\text{MC}$) as well as on the SB of Pythia data ($\delta_\text{sys}^\text{SB}$). We also show a quadratic addition of the Herwig MC and SB estimates ($\delta_\text{sys}^{\text{MC}\oplus\text{SB}}$) as described in the text. The total statistical error $\sigma_\text{stat}$ defined in eqn.~(\ref{eq:sigma stat}) is also shown in the lower panel to guide the eye as to how relevant the observed deviations between the different estimates of $\delta_\text{sys}$ are. Note that $\sigma_\text{stat}$ is not an error on $\delta_\text{sys}$. The results are based on 10 classifier runs in each signal region run on independent density estimation samples.}
    \label{fig:Rsys cathode}
\end{figure}

For \Cathode, the estimates for $\delta_\text{sys}$ are again shown for our Herwig MC ($\delta_\text{sys}^\text{MC}$) as well as our (Pythia) data ($\delta_\text{sys}^\text{data}$) in Figure~\ref{fig:Rsys cathode} and in Table~\ref{tab: Rsys all}. Note the reduced statistical error for \Cathode{} due to oversampling. For $\epsilon_B=0.001$ or larger, the MC estimate is again reliable. However, for smaller $\epsilon_B \approx 0.0001$, the MC clearly underestimates $\delta_\text{sys}$: $\delta_\text{sys}^\text{data} \approx 0.7$ is much larger than  $\delta_\text{sys}^\text{MC} \approx 0.25$. 
Evidently, the data (Pythia) probability distribution seems to be more difficult to model in the tails than the MC (Herwig) probability distribution. This is probably a general pitfall of using a combination of generative modeling and MC to estimate $\delta_{\rm sys}$: any data/simulation differences are likely to be exacerbated by the generative model on the tails of probability distributions. (By contrast, $\delta_{\rm sys}$ was accurately estimated even on the tails by Herwig in the case of CWoLa, where no generative model was involved.)

In Section~\ref{sec:deltasysSB}, we have introduced an alternative data-driven method for estimating $\delta_\text{sys}=\delta_\text{sys}^\text{SB}$. In Figure~\ref{fig:Rsys cathode} we compare the results of this data-driven method with $\delta_\text{sys}^\text{data}$. We present $\delta_\text{sys}^\text{SB}$ based on side-band data without signal ($S/B =0$). $\delta_\text{sys}^\text{SB}$ with the default signal injection ($S/B\approx0.64\%$) is shown in Figure~\ref{fig:Rsys cathode with signal} in Appendix~\ref{app:Rsys}. 

The data-driven estimate $\delta_\text{sys}^\text{SB}$ is not very sensitive to potential signal contamination or to the choice of feature set. For $\epsilon_B\gtrsim 0.001$ we slightly underestimate $\delta_\text{sys}$ because the interpolation error is not taken into account. In this region the Monte Carlo estimate is valuable. However, for $\epsilon_B \approx 0.0001$, where density estimation itself is difficult for the Pythia data and $\delta_\text{sys}^\text{MC}$ underestimates $\delta_\text{sys}$, we get a reliable data-driven estimate for $\delta_\text{sys}$. Hence, the two methods show a nice complementarity since they focus on different failure modes of estimating $\delta_\text{sys}$.

\begin{figure}[t]
    \centering
    \subfloat[\Cathode: $S/B=0\%$]{\includegraphics[width=0.49\textwidth]{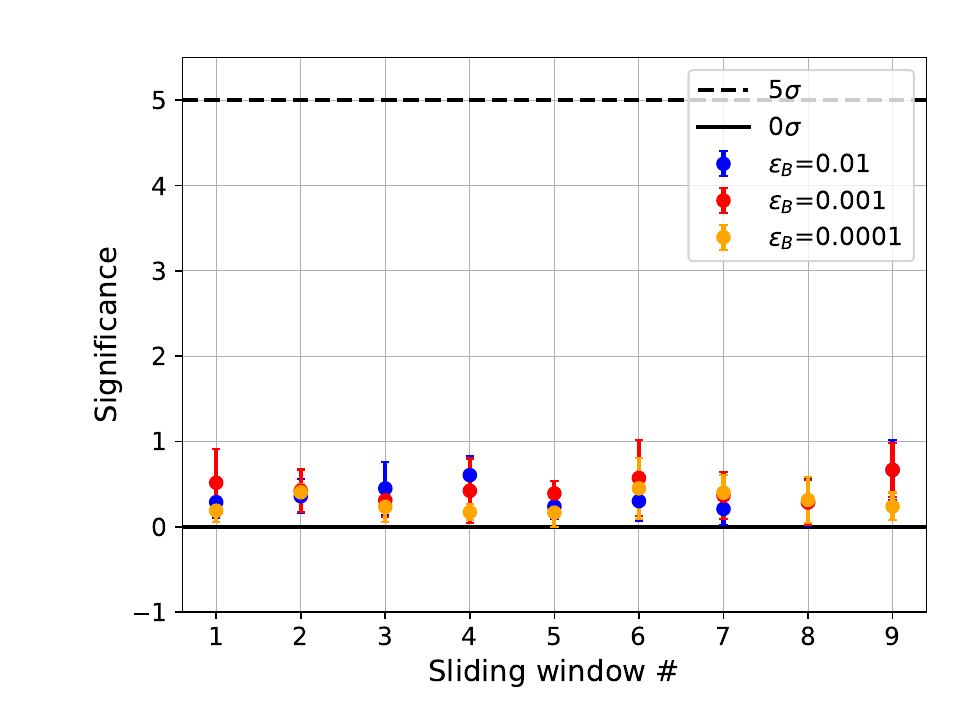}}
    \subfloat[\Cathode\ $\Delta R$: $S/B=0\%$]{\includegraphics[width=0.49\textwidth]{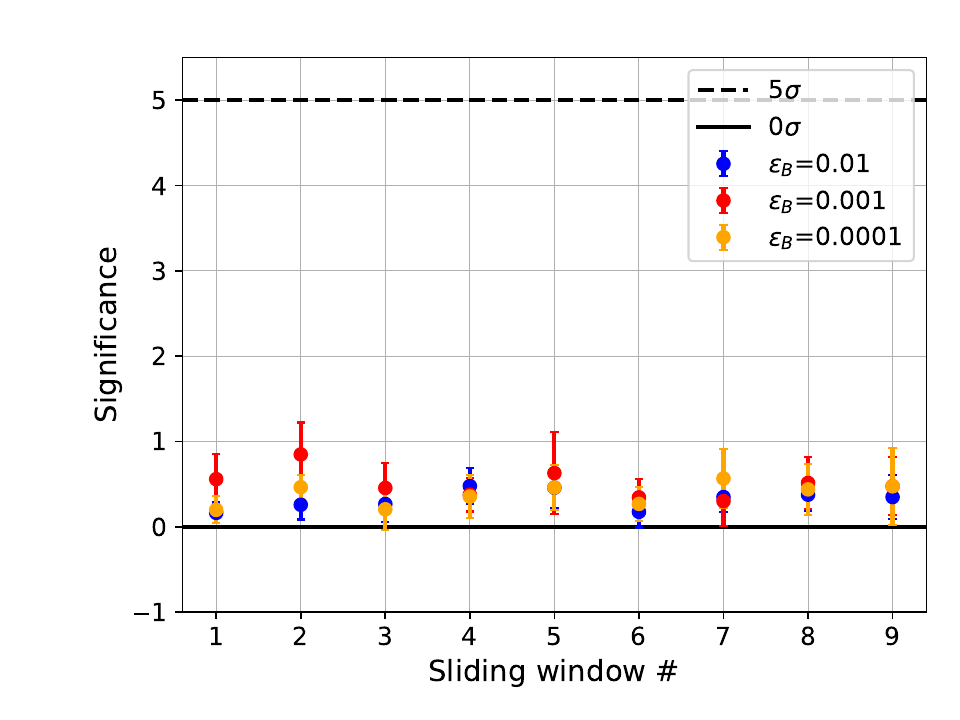}}\\
    \subfloat[\Cathode: $S/B=0.64\%$]{\includegraphics[width=0.49\textwidth]{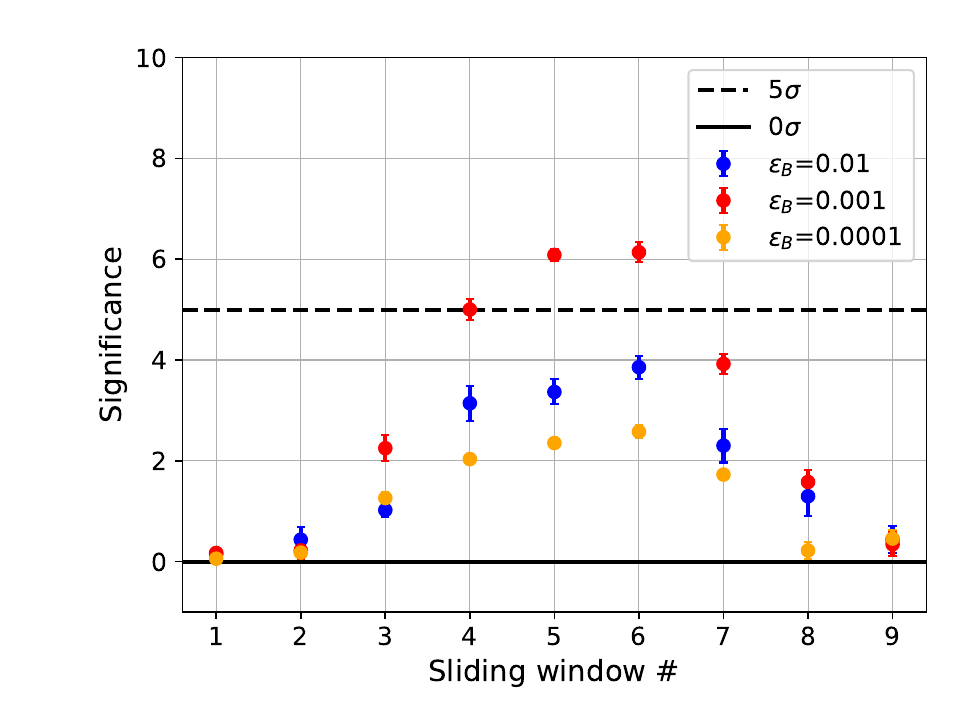}} 
    \subfloat[\Cathode\ $\Delta R$: $S/B=0.64\%$]{\includegraphics[width=0.49\textwidth]{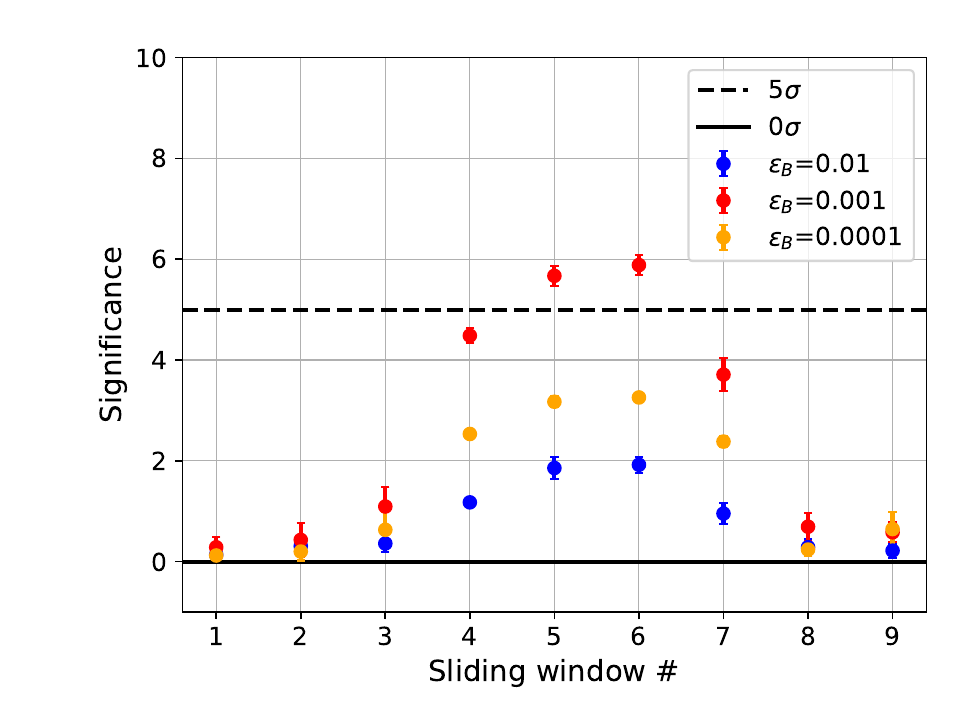}}
    \caption{Significance ${\cal S}$, eqn.~(\ref{eq:significance}), for the different signal regions for \Cathode\ without (top) and with signal injection (bottom) using the baseline dataset (left) and the dataset with $\Delta R$ (right). The error bars indicate the variance of the significance based on 10 classifier runs.}
    \label{fig:significances cathode}
\end{figure}

For the results presented in Figure~\ref{fig:significances cathode}, we again use $\sigma_\text{sys}=\delta_\text{sys}$, where now we use the more robust choice $\delta_\text{sys}=\delta^{\text{MC}\oplus\text{SB}}_\text{sys}=\sqrt{\left(\delta^{\text{MC}}_\text{sys}\right)^2+\left(\delta^{\text{SB}}_\text{sys}\right)^2}$. 
In panels (a) and (b), we see the significance on a data set without signal, which is in good agreement with the null hypothesis, in particular also for $\epsilon_B=0.0001$. Panels (c) and (d) show the significance when the signal is present. Here we observe a significant deviation from the null hypothesis, especially for the working point $\epsilon_B=0.001$. For $\epsilon_B=0.0001$ the significance is reduced due to the more conservative (data driven) error estimate. Contrary to the CWoLa method, the addition of $\Delta R$ does not reduce the significance for \Cathode.  

\section{Conclusion}
\label{sec:Conclusion}

In this paper we demonstrate how resonant anomaly detection methods can transform a traditional bump hunt into a straightforward cut-and-count experiment: We cut on the anomaly score and compare the observed number of data events with the expected number of background in the absence of anomalous events. By doing so, we eliminate the common problem of background sculpting that arises when using resonant anomaly detection in traditional fit-based bump hunts. Furthermore, our approach allows for large background rejection rates, where weakly supervised methods typically perform best.

In our cut-and-count approach, we need to estimate the systematic bias caused by an imperfect background template. This is accomplished by introducing the systematic shift $\delta_\text{sys}$ in Section~\ref{sec:Sensitivity}. We estimate $\delta_\text{sys}$ on background Monte Carlo simulation and, for CATHODE, also in a data-driven manner. We quantify the performance of this method on the LHC Olympics R\&D dataset, which includes a heavy resonance decaying into jets as a new physics signal benchmark. We use the CWoLa and CATHODE methods to define the background template, using a small set of high-level observables as input features, together with a powerful classifier based on a boosted decision tree. We observe no false discoveries in the background-only case and are able to detect a signal injected at approximately $2\,\sigma$ beyond $5\,\sigma$ for both CWoLa and CATHODE on our baseline dataset. 

When we deliberately break the assumptions of CWoLa by adding a feature correlated with the dijet mass to the dataset, we still do not observe any false discoveries as $\delta_\text{sys}$ increases dramatically. This highlights an interesting feature of $\delta_\text{sys}$, which is its potential as a signal-independent method for assessing the quality of the background template. We observe this same feature for CATHODE in the tails of the distribution, which are difficult to estimate with a density estimator, something that is clearly reflected in the behavior of $\delta_\text{sys}$ at large background rejections.

Our estimates for $\delta_\text{sys}$ and the corresponding systematic error $\sigma_\text{sys}$ are rather straightforward. Using Monte Carlo simulation, the systematic effects could be further explored and significantly refined. For example, using multiple simulated data sets, $\delta_\text{sys}$ could be estimated individually for each signal region using the mean across different data realizations. Additionally, the sensitivity to differences in the Monte Carlo modeling could be studied by using a broader array of simulation tools which allow for a better estimate of $\sigma_\text{sys}$. 
A more elaborate analysis is left for future studies. 

When considering the final significances 
it is important to note that we have chosen a large systematic error $\sigma_\text{sys}=\delta_\text{sys}$, which naturally reduces the significance with which we detect the signal. Our estimates of $\delta_\text{sys}$ are likely to be more reliable than this error suggests (see Figures \ref{fig:Rsys cwola} and \ref{fig:Rsys cathode}). Further studies using independent Monte Carlo datasets may provide deeper insights into the behaviour of $\delta_\text{sys}$ and thus allow for a reduction of $\sigma_\text{sys}$.

On a data set with a broader invariant mass spectrum, a data-driven validation of $\delta_\text{sys}$ would also be possible: As we assume a signal localized in $m_{JJ}$, the majority of the spectrum should be signal free and therefore, provided a good estimate of $\delta_\text{sys}$, compatible with the null hypothesis. Therefore, if deviations are observed across the whole spectrum, the estimate of $\delta_\text{sys}$ and the analysis results should be reconsidered.

In this proof-of-concept study, we have examined a small set of high-level observables. However, our method is not limited to this particular case and can be easily generalized to larger feature sets and also to non-resonant anomaly searches~\cite{Bickendorf:2023nej, Bai:2023yyy, Kasieczka:2024lxf, Finke:2022lsu}, provided that powerful classification algorithms and methods for obtaining a background template are available. Such studies are reserved for future work.

\section*{Acknowledgements}
MH is supported by the Deutsche Forschungsgemeinschaft (DFG) under grant 400140256 -- GRK 2497: Physics of the heaviest particles at the Large Hadron Collider. The research of MK and AM is supported by the DFG under grant 396021762 -- TRR 257: Particle physics phenomenology after the Higgs discovery. GK acknowledges support by the DFG under the German Excellence Initiative -- EXC 2121  Quantum Universe – 390833306. The work of RD and DS is supported by U.S.\ Department of Energy grant DE-SC0010008. This research used resources provided  by RWTH Aachen University under project rwth0934 and by the National Energy Research Scientific Computing Center, a DOE Office of Science User Facility supported by the Office of Science of the U.S.\ Department of Energy under Contract No. DE-AC02-05CH11231 using NERSC award HEP-ERCAP0027491. The authors thank the Office of Advanced Research Computing (OARC) at Rutgers, The State University of New Jersey \url{https://it.rutgers.edu/oarc} for providing access to the Amarel cluster and associated research computing resources that contributed to the results reported here. This work was performed in part at the Aspen Center for Physics, supported by National Science Foundation grant PHY-2210452.

\section*{Code}
The code for this paper can be found at \href{https://github.com/mariehein/bumphunt_paper/tree/main}{\texttt{https://github.com/mariehein/bumphunt\_paper /tree/main}}.

\clearpage

\appendix
\section{Architecture and training}
\label{app:Architecture}
\subsection{The boosted decision tree classifier}\label{app:bdt}
As in \cite{Finke:2023ltw}, we use the \texttt{HistGradientBoostingClassifier} from \texttt{scikit-learn}. \cite{Pedregosa:2011sk}, which is based on LightGBM \cite{Ke:2017lgbm}. It is a gradient boosted decision tree (BDT) that achieves high training and evaluation speed by histogramming its input features. We largely use default hyperparameters, such as a learning rate of 0.1, a maximum number of leaf nodes per tree of 31, and a maximum number of bins per feature of 255. We also use early stopping with a patience of 10 iterations. The maximum number of iterations was increased to 200, which is rarely used, but was done to ensure that early stopping and not the maximum number of iterations leads to the end of training. 

What we call a classifier is an ensemble of 50 such BDTs with randomized training and validation splits. This was found in \cite{Finke:2023ltw} to give stable and good performance on a variety of datasets without further hyperparameter tuning. 

\subsection{Density estimation with Conditional Flow Matching}\label{app:flow} 
Conditional Flow matching (CFM) is a faster and more feasible way to train Continuous Normalizing Flows (CNFs) \cite{cnf}. In CNF, one attempts to learn the vector field $u_t(x_t): [0,1] \times \mathbb{R}^d \to \mathbb{R}^d$, which generates a continuous transformation of data $x_t$:

\begin{equation}
    \frac{d x_t}{dt} = u_t(x_t),
\end{equation}
where at $t=0$, $x_0$ follows the data distribution $p_\text{data}(x_0)$, and at $t=1$, $x_1$ follows a known distribution $p_{\text{base}}(x_1)$. We use the normal distribution $\mathcal{N}(x|0,I)^d$ as $p_{\text{base}}(x)$. For general $t$, the $x_t$ generated by the vector field follows a density $p_t(x_t)$. A CNF trained by maximizing the likelihood  drastically increases the computational cost, since evaluating the likelihood requires solving an ODE for each data point and model iteration.

The key idea in Conditional Flow Matching (CFM) is to learn the conditional vector field $u_t(x_t|x_0)$ which generates a conditional probability path $p_t(x|x_0)$. At $t=0$, we have $p_0(x|x_0)=\mathcal{N}(x|x_0,\sigma^2 I
)$ where $\sigma^2$ is very small, whereas at $t=1$, we have $p_1(x|x_0) = \mathcal{N}(x|0,I)^d.$ Marginalizing this conditional density over $p_{\rm data}(x_0)$ gives us the unconditional probability $p_t(x)$:
\begin{equation}
p_t(x)=\int d x_0 p_t(x|x_0) p_{\rm data}(x_0).
\end{equation}
 In CFM, this conditional vector field is regressed with a neural network $v_{\theta}(x_t|t)$ by minimizing the CFM loss
\begin{equation}
\mathcal{L}(\theta) = \left\|v_{\theta}(x_t|t)-u_t\left(x_t|x_0\right)\right\|^2,
\end{equation}
which is averaged over $t \sim \mathcal{U}[0,1]$, $x_0 \sim p_{\text{data}}(x_0)$ and $x_t \sim  p_t(x|x_0)$. The authors in \cite{lipman2023flowmatchinggenerativemodeling} show that by learning $u_t(x_t|x_0),$ one also learns $v_{\theta}(x|t) = u_t(x)$. Aside from $t$, our models also have $m$, the resonant feature, as a conditional feature which allows us to model the vector field $u_t(x|m)$ corresponding to $p(x|m)$. 

We use the same \texttt{ResNet}-style \cite{resnet} architecture from \texttt{nflows}\cite{nflows} that was used to model $v^{\rm {CR}}_{\theta}$ described in section II.C.2 of \cite{sigma}. Similarly, the model and training hyperparameters are the same as Section II.D of \cite{sigma}.

\section{Further studies of the systematic shift $\delta_\text{sys}$}
\label{app:Rsys}

\subsection{Systematic shift $\delta_\text{sys}$ for CATHODE in the presence of signal}
\begin{figure}[t]
    \centering
    \subfloat[CATHODE Baseline]{\includegraphics[width=0.49\textwidth]{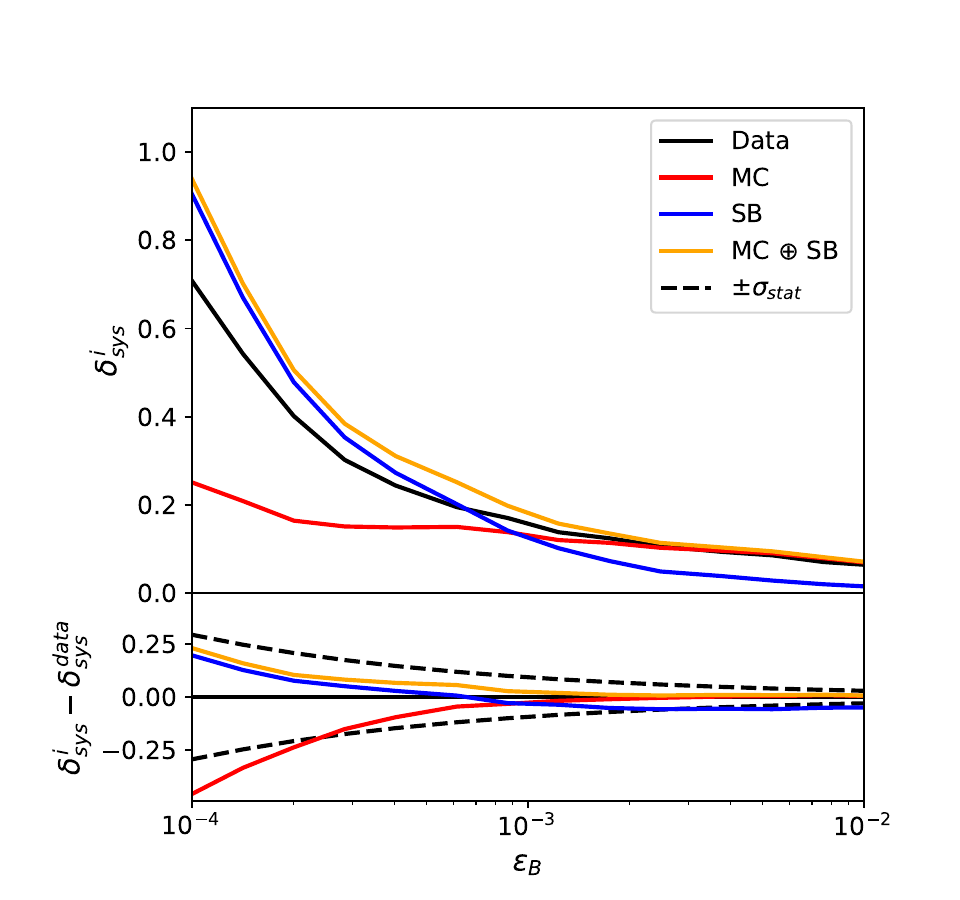}}  
    \subfloat[CATHODE $\Delta R$]{\includegraphics[width=0.49\textwidth]{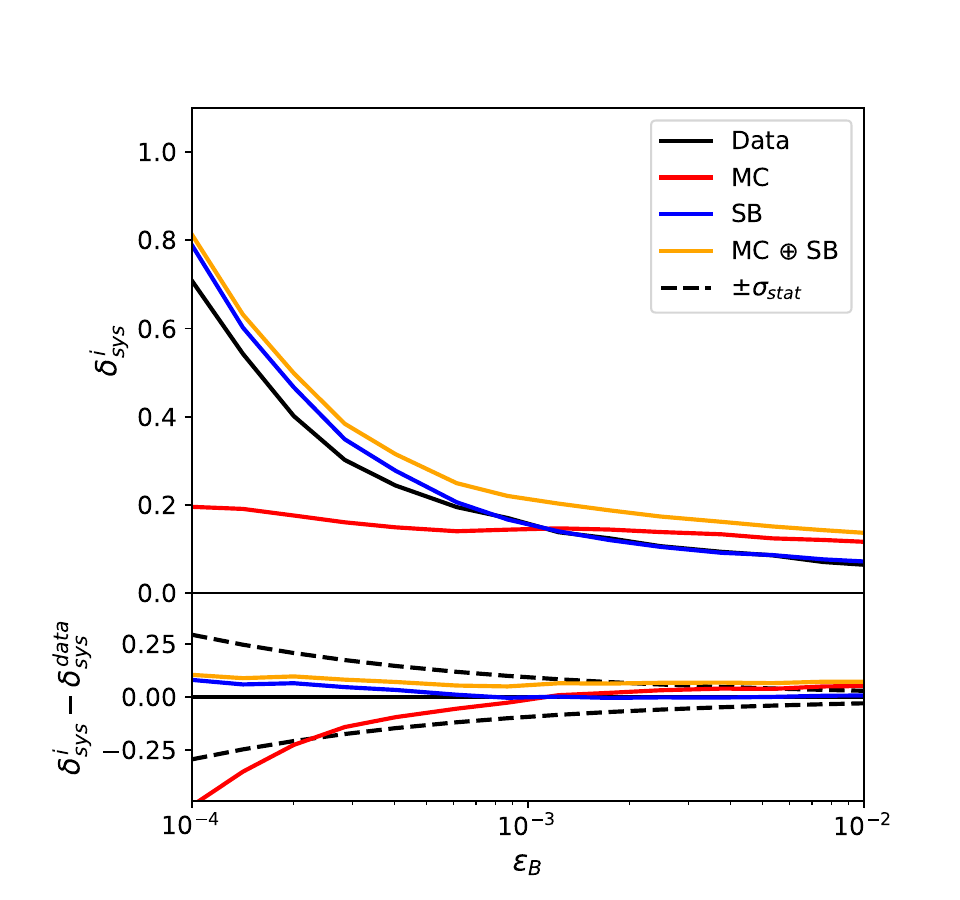}}\\
    \caption{Same as Figure \ref{fig:Rsys cathode} but for the SB estimate of $\delta_\text{sys}$ the data set contains signal ($S/B\approx0.64\%$).}
    \label{fig:Rsys cathode with signal}
\end{figure}

In Figure \ref{fig:Rsys cathode}, we showed different estimates of $\delta_\text{sys}$ obtained for CATHODE. In the case of the data-driven estimate $\delta_\text{sys}^\text{SB}$ -- and therefore also the combined estimate $\delta_\text{sys}^{\text{MC}\oplus\text{SB}}$ -- this estimation is affected by the presence of signal in the data set. Therefore, in Figure \ref{fig:Rsys cathode with signal}, we show the same plots as in Figure \ref{fig:Rsys cathode} but include signal in the data set, on which $\delta_\text{sys}^\text{SB}$ is estimated.

Comparing both figures, we see a slightly enlarged value of $\delta_\text{sys}^\text{SB}$ in the presence of signal as the classifier can identify the small number of signal events in the data set. $\delta_\text{sys}^{\text{MC}\oplus\text{SB}}$ is affected accordingly. For the analysis, this means that significances are reduced slightly when signal is present compared to when it is not. However, by using the whole sideband and therefore diluting the signal for this analysis as was described in Section \ref{sec:systematic_shift_and_error}, this effect is minimal.

\subsection{Systematic shift $\delta_\text{sys}$ at the different working points}
Table \ref{tab: Rsys all} contains all values of $\delta_\text{sys}$ and $\sigma_\text{stat}$ at the three working points used for the significance plots. For CWoLa and Cathode the information on $\delta_\text{sys}$ is already contained in Figures \ref{fig:Rsys cwola}, \ref{fig:Rsys cathode}, and \ref{fig:Rsys cathode with signal}.

\begin{table}[t]
    \centering
    \begin{tabular}{|c|c|c|c|c|c|c|c|}\hline
         \multicolumn{2}{|c|}{Data set} & \multicolumn{3}{c|}{Baseline} & \multicolumn{3}{c|}{$\Delta R$}\\\hline\hline
         \multicolumn{2}{|c|}{$\epsilon_B$} & 0.01 & 0.001 & 0.0001 & 0.01 & 0.001 & 0.0001\\\hline\hline
         \multirow{2}{*}{IAD} &$\delta_\text{sys}^\text{data}$ & -0.01 & 0.01 & 0.05 & 0.01 & 0.06 & 0.20 \\\cline{2-8}
         &$\sigma_\text{stat}$ & 0.04 & 0.13 & 0.34 & 0.04 & 0.12 & 0.34 \\\hline
         \multirow{3}{*}{CWoLa}&$\delta_\text{sys}^\text{data}$ & 0.14 & 0.16 & 0.20 & 1.01 & 1.02 & 1.02 \\\cline{2-8}
         &$\delta_\text{sys}^\text{MC}$ & 0.14 & 0.11 & 0.05 & 0.98 &  1.04 & 0.94 \\\cline{2-8}
         &$\sigma_\text{stat}$ & 0.04 & 0.13 & 0.35 & 0.04 & 0.13 & 0.35 \\\hline
         \multirow{5}{*}{CATHODE}&$\delta_\text{sys}^\text{data}$ & 0.06 & 0.16 & 0.71 & 0.13 & 0.22 & 0.50 \\\cline{2-8}
         &$\delta_\text{sys}^\text{MC}$ & 0.07 & 0.14 & 0.25 & 0.12 & 0.15 & 0.20 \\\cline{2-8}
         &$\delta_\text{sys}^\text{SB}(S/B=0)$ & 0.01 & 0.08 & 0.77 & 0.06 & 0.10 & 0.66 \\\cline{2-8}
         &$\delta_\text{sys}^{\text{MC}\oplus\text{SB}}(S/B=0)$ & 0.07 & 0.16 & 0.81 & 0.13 & 0.18 & 0.68 \\\cline{2-8}
         &$\delta_\text{sys}^\text{SB}(S/B=0.64\%)$ & 0.02 & 0.13 & 0.90 & 0.07 & 0.16 & 0.79 \\\cline{2-8}
         &$\delta_\text{sys}^{\text{MC}\oplus\text{SB}}(S/B=0.64\%)$ & 0.07 & 0.19 & 0.94 & 0.14 & 0.21 & 0.81 \\\cline{2-8}
         &$\sigma_\text{stat}$ & 0.03 & 0.09 & 0.30 & 0.03 & 0.09 & 0.30 \\\hline
    \end{tabular}
    \caption{Relative systematic shift $\delta_\text{sys}$ estimated from an analysis without signal on Pythia data ($\delta_\text{sys}^\text{data}$), Herwig MC ($\delta_\text{sys}^\text{MC}$) as well as for CATHODE in a data-driven manner on the SB of Pythia data with and without signal ($\delta_\text{sys}^\text{SB}(S/B=0)$ and $\delta_\text{sys}^\text{SB}(S/B=0.64\%)$ respectively). For the IAD $\delta_\text{sys}^\text{data}$ is determined on the Pythia reproduction of the LHCO R\&D data set. The results are based on 10 classifier runs in each signal region. For Cathode, independent DEs are used for each classifier run. For reference, the statistical error $\sigma_\text{stat}$ is also given.}
    \label{tab: Rsys all}
\end{table}

\subsection{Dependence of $\delta_\text{sys}$ on the signal region window}

Throughout this work, we assume that for each working point $\epsilon_B$ the relative systematic shift $\delta_\text{sys}$ is constant across all signal regions. The validity of this choice will be discussed in this section.

\begin{figure}[t]
    \centering
    \subfloat[CWoLa Data]{\includegraphics[width=0.49\textwidth]{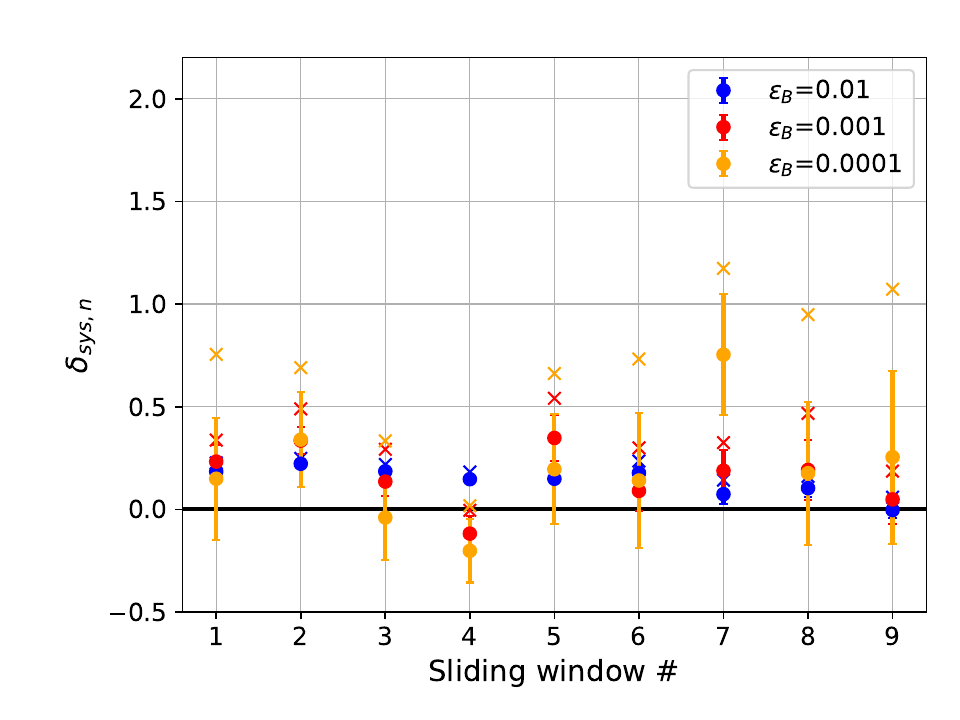}}
    \subfloat[CWoLa MC]{\includegraphics[width=0.49\textwidth]{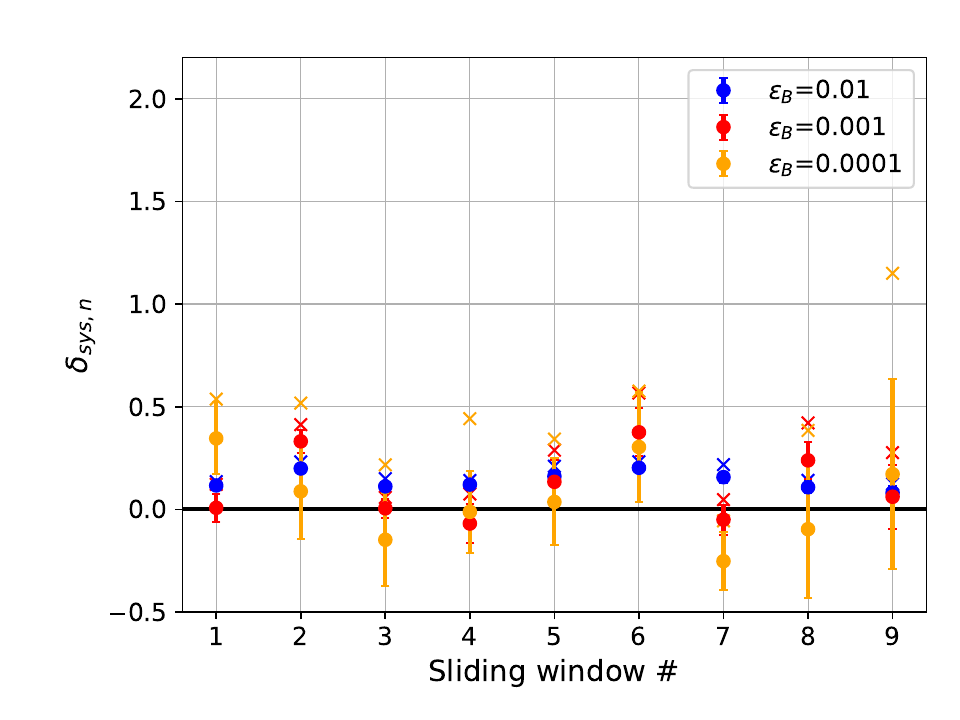}}
    \caption{Relative systematic shift $\delta_\text{sys,n}$ (see Equation \ref{eq:R}) per window for CWoLa determined on Pythia data (left) and Herwig MC (right) for the baseline dataset. The error bars indicate the variance based on 10 classifier runs, the $\times$ indicates the maximum observed value.}
    \label{fig: R per window cwola}
\end{figure}

Figure \ref{fig: R per window cwola}, panels (a) and (b), show the values of $\delta_{\text{sys},n}$ obtained for CWoLa on the different windows for the baseline feature set. There is no significant dependence of $\delta_{\text{sys},n}$ on the window number, so choosing a constant value of $\delta_{\text{sys},n}$ across the windows is a reasonable approximation.

\begin{figure}[t]
    \centering
    \subfloat[\Cathode\ Data]{\includegraphics[width=0.49\textwidth]{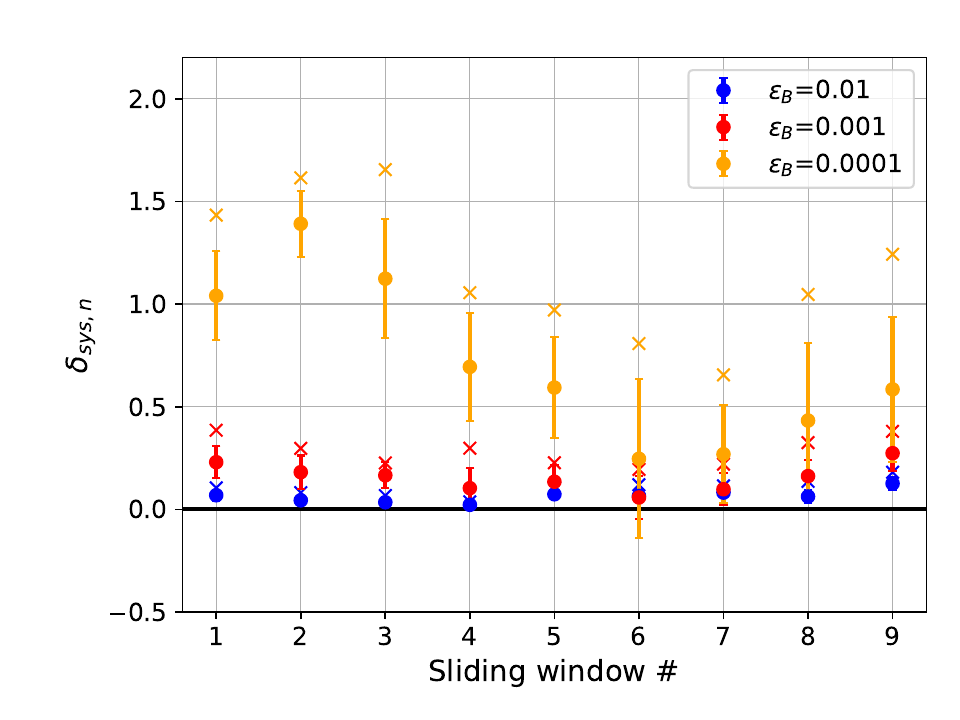}}
    \subfloat[\Cathode\ MC]{\includegraphics[width=0.49\textwidth]{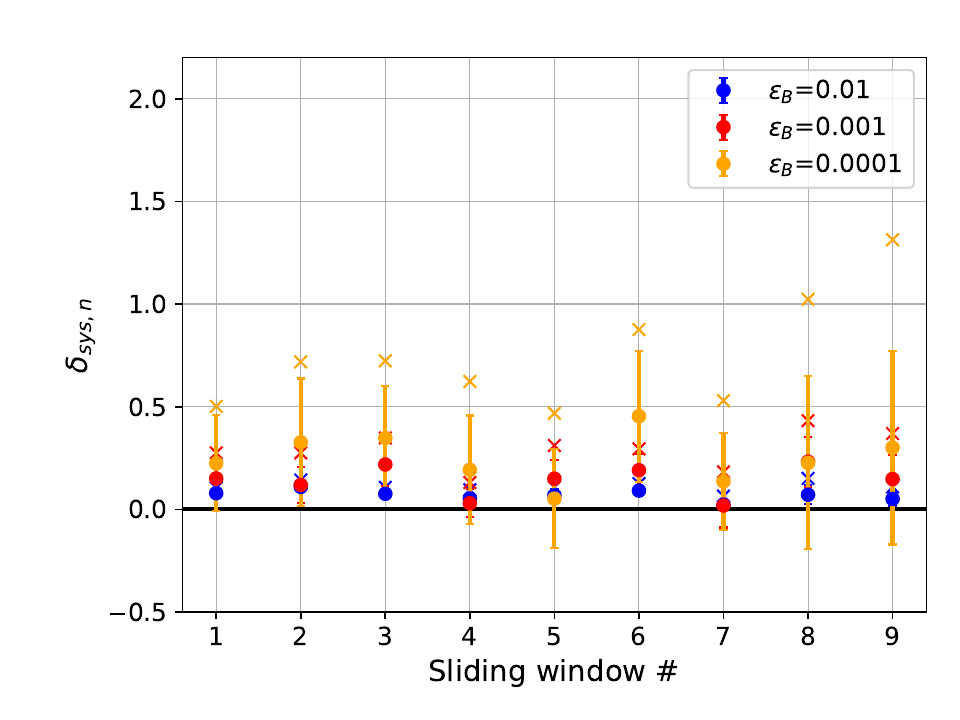}}\\
    \subfloat[\Cathode\ SB, $S/B=0$]{\includegraphics[width=0.49\textwidth]{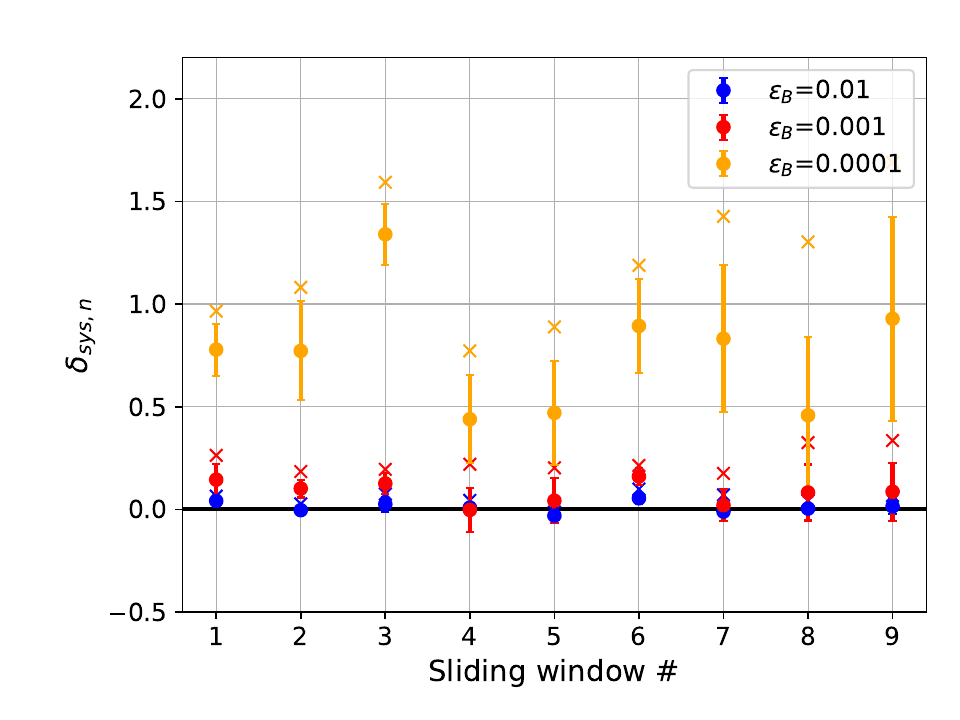}}
    \subfloat[\Cathode\ SB, $S/B=0.64\%$]{\includegraphics[width=0.49\textwidth]{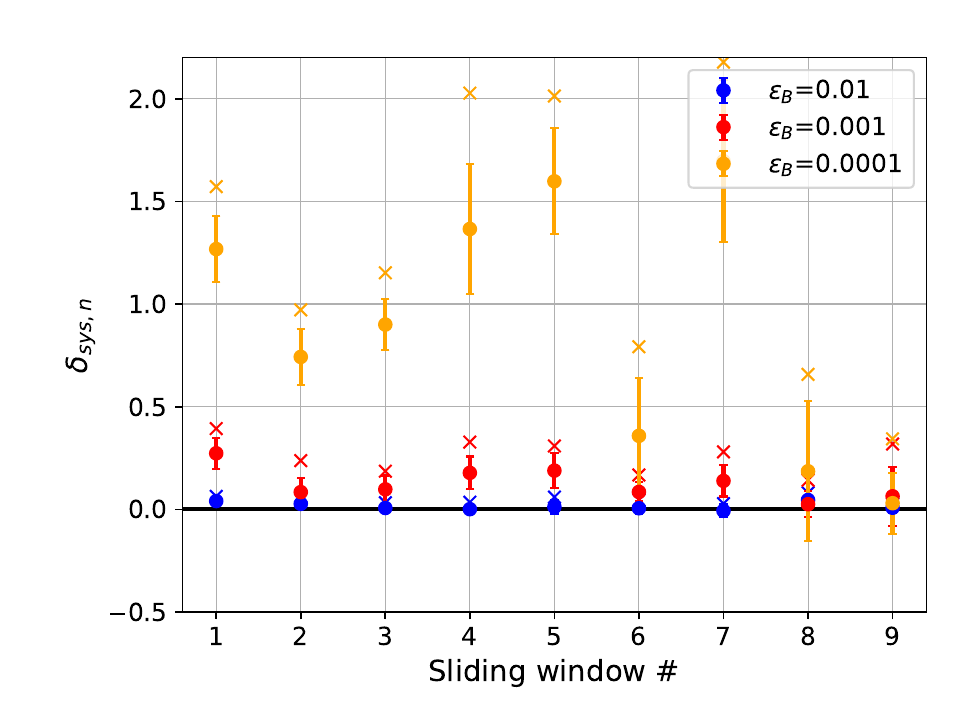}}
    \caption{Relative systematic shift $\delta_\text{sys,n}$ (see Equation \ref{eq:R}) per window for Cathode determined on Pythia data (top left), Herwig MC (top right) and on Pythia data using the SB (bottom) without (left) and with (right) signal for the baseline dataset. The error bars indicate the variance based on 10 classifier runs, the $\times$ indicates the maximum observed value.}   \label{fig: R per window cathode}
\end{figure}

For Cathode, more structure is visible in Figure \ref{fig: R per window cathode}, at least at the lowest working point of $\epsilon_B=0.0001$. If we focus on the true distribution, Figure \ref{fig: R per window cathode}(a), we see particularly large values of $\delta_{\text{sys},n}$ in windows one to three. Since these windows are closest to the trigger turn-on, it is possible that this is where the shape originates.
However, a conclusive answer to this question would require further study.

The shape observed for $\delta_{\text{sys},n}$ on data is not seen in the MC estimation, Figure \ref{fig: R per window cathode}(b), which is very smooth in general. Neither is it seen in the data-driven estimation on the sidebands without signal, Figure \ref{fig: R per window cathode}(c), where $\delta_{\text{sys},n}$ seems to be more constant with some statistical fluctuation. Therefore, a window-by-window estimation of $\delta_\text{sys}$ would not be able to decrease mismodeling. 

The data-driven estimation on the sidebands with signal, Figure \ref{fig: R per window cathode}(d), shows larger window to window fluctuations, which do not seem to significantly impact the variance observed in each window.

\subsection{Dependence of $\delta_\text{sys}$ on the background template quality}

To study the effect of the background template quality on $\delta_\text{sys}$ and, therefore, on the analysis as a whole, we compare the $\delta_\text{sys}$ values obtained using the samples from flow matching (as described in Appendix \ref{app:flow}) to the samples from MAF (used in Ref. \cite{Hallin:2021wme}). Table \ref{tab:Rsys old new DE} shows $\delta_\text{sys}$ for both the density estimators. For both data sets, we see that $\delta_\text{sys}$ is lower at every working point for the flow matching than for the MAF. This is also observed for $\delta_\text{sys}^\text{MC}$ and $\delta_\text{sys}^\text{SB}$. This lower $\delta_\text{sys}$ using the samples from flow matching also results in a higher discovery significance in an analysis with signal.

\begin{table}[t]
\centering
\begin{tabular}{|c|c|c|c|c|c|c|}\hline
     Data set & \multicolumn{3}{c|}{Baseline} & \multicolumn{3}{c|}{$\Delta R$}\\\hline\hline
     $\epsilon_B$ & 0.01 & 0.001 & 0.0001 & 0.01 & 0.001 & 0.0001\\\hline\hline
     Flow Matching & 0.06 & 0.16 & 0.71 & 0.13 & 0.22 & 0.50 \\\hline
     MAF & 0.13 & 0.25 & 0.80 & 0.20 & 0.28 & 0.56 \\\hline
     $\sigma_\text{stat}$ & 0.03 & 0.09 & 0.30 & 0.03 & 0.09 & 0.30 \\\hline
     \end{tabular}
    \caption{Relative systematic error $\delta_\text{sys}^\text{data}$ estimated from an analysis  without signal on Pythia data. The results are based on 10 classifier runs in each signal region using samples from independent DE trainings. For reference, the statistical error $\sigma_\text{stat}$ is also given.}
    \label{tab:Rsys old new DE}
\end{table}

\section{Further studies of the observed significances}
\label{app:further_studies}

\subsection{Comparing observed IAD significances to SIC values}\label{sec:significances IAD}

The naive significance improvement characteristic (SIC) value, ${\rm SIC} = \epsilon_S/\sqrt{\epsilon_B}$, which is often reported to quantify the anomaly detection potential, is approximately 11 for the working point $\epsilon_B=10^{-3}$ point~\cite{Finke:2023ltw}. 
Thus, with an initial significance of 2.2, the naively expected significance is about 24. For our IAD analysis, $\sigma_\text{exp}$ is equal to the statistical error because we estimate $\epsilon_B$ on a background template of the same size as the data set. Thus, using the formula for the Gaussian limit eqn.~(\ref{eq:significance_Gaussian_limit}) results in a significance of about $\mathcal{S}_G\approx24/\sqrt{2}\approx17$. (This loss of performance can be avoided by using oversampling, as is possible for \Cathode.) Using instead the proper Poisson statistics for $N_\text{exp}\approx 130$ at this working point by employing eqn.~(\ref{eq:significance}), the significance is finally reduced to $\mathcal{S}=12$. The difference to $\mathcal{S}=8$ in our cut and count analysis is due to the fact that, unlike Ref.~\cite{Finke:2023ltw}, we do not use an oversampled background template and the k-fold cross validation does not use an independent test set. Furthermore, the random but fixed signal sample used in Ref.~\cite{Finke:2023ltw} and throughout our analysis is particularly difficult to classify using k-fold cross validation. Drawing a different random signal sample generally leads to a higher significance.  

\subsection{Significances for CATHODE using $\delta_\text{sys}^\text{MC}$}

\begin{figure}[t]
    \centering
    \subfloat[\Cathode: $S/B=0\%$]{\includegraphics[width=0.49\textwidth]{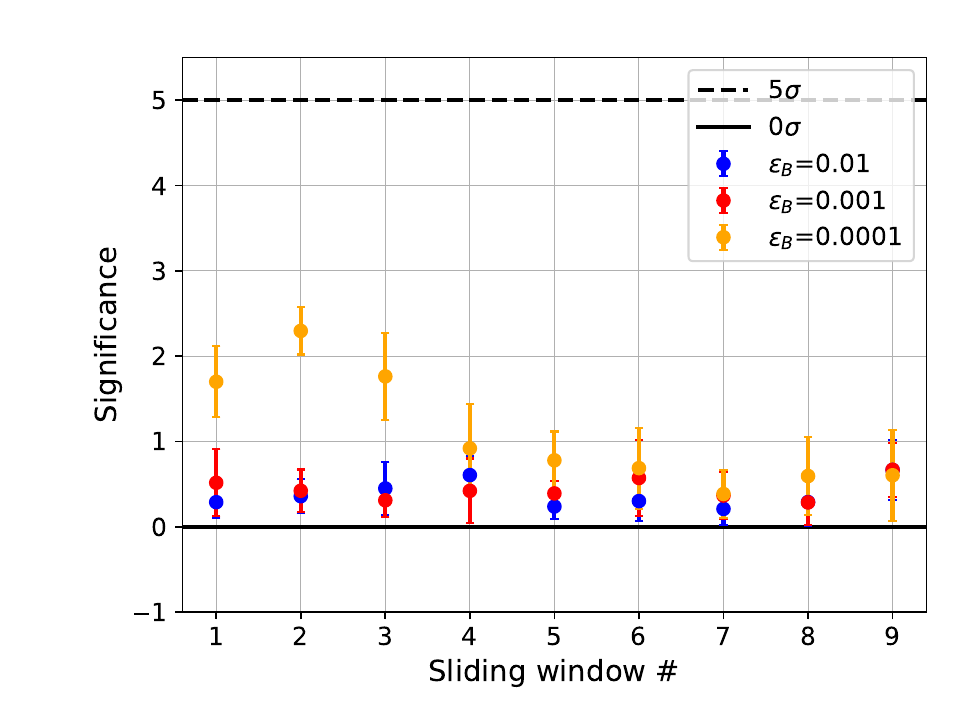}}
    \subfloat[\Cathode\ $\Delta R$: $S/B=0\%$]{\includegraphics[width=0.49\textwidth]{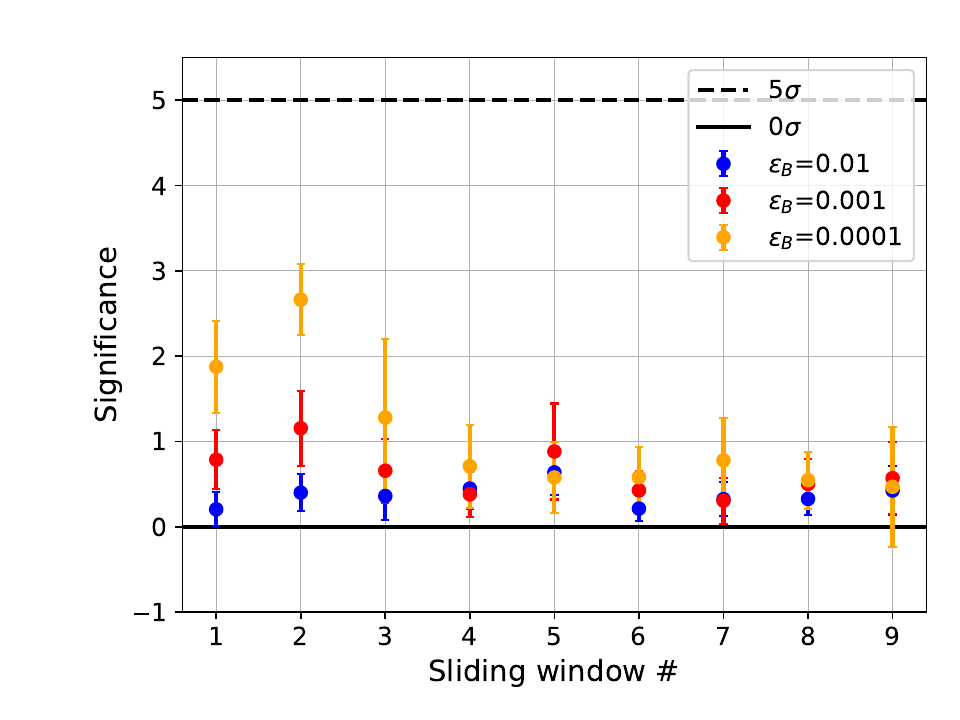}}\\
    \subfloat[\Cathode: $S/B=0.64\%$]{\includegraphics[width=0.49\textwidth]{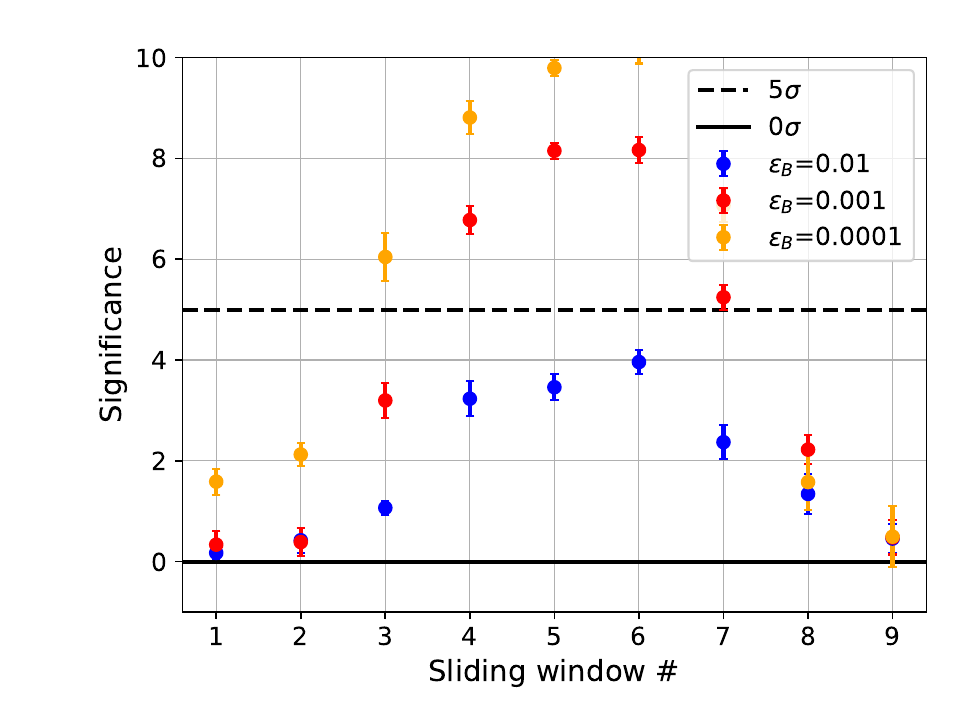}} 
    \subfloat[\Cathode\ $\Delta R$: $S/B=0.64\%$]{\includegraphics[width=0.49\textwidth]{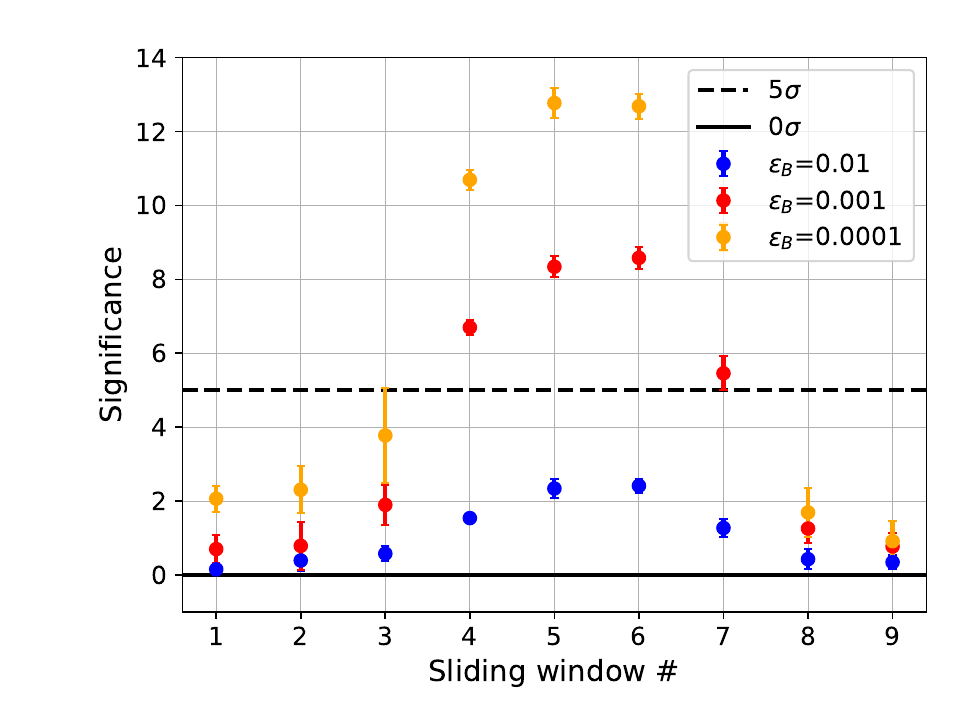}}
    \caption{Significance ${\cal S}$, eqn.~(\ref{eq:significance}), for the different signal regions for \Cathode{} using  $\sigma_\text{sys}=\delta_\text{sys}=\delta_\text{sys}^\text{MC}$ without (top) and with signal injection (bottom) using the baseline dataset (left) and the dataset with $\Delta R$ (right). The error bars indicate the variance of the significance based on 10 classifier runs.}
    \label{fig:significances cathode original}
\end{figure}

For CATHODE, we use $\delta_\text{sys}^{\text{MC}\oplus\text{SB}}$ in Section \ref{subsec:Cathode} at the systematic shift as this value corresponds very well with the reference value $\delta_\text{sys}^\text{data}$ across all working points. Since this choice does deviate from the choice made for CWoLa, where we use $\delta_\text{sys}^\text{MC}$, we show the significances obtained by using $\delta_\text{sys}^\text{MC}$ for CATHODE in Figure \ref{fig:significances cathode original}. 
In the signal-free case, panels (a) and (b), we see a good agreement with the null hypothesis for $\epsilon_B=10^{-2}$ and $10^{-3}$, where $\delta_\text{sys}^{\text{MC}\oplus\text{SB}}$ is dominated by $\delta_\text{sys}^\text{MC}$. For $\epsilon_B=10^{-4}$ on the other hand, we observe a spurious peak around signal window 2. Nevertheless, the significance is still below $3\sigma$ here despite the significant underestimation of $\delta_\text{sys}$.

With signal, panels (c) and (d), the low systematic shift results in a high significance at the same threshold. Comparing with Figure \ref{fig:significances cathode}, we obtain a higher significance for $\epsilon_B=10^{-3}$ as well as both contributions to $\delta_\text{sys}^{\text{MC}\oplus\text{SB}}$ are of similar size here. For $\epsilon_B=10^{-2}$, we remain below $5\sigma$ as here $\delta_\text{sys}^{\text{MC}\oplus\text{SB}}$ is dominated by $\delta_\text{sys}^\text{MC}$.

\clearpage

\bibliographystyle{JHEP_improved}
\bibliography{bibliography.bib}

\end{document}